\documentclass[aps,pra,showpacs]{revtex4}
\usepackage{bbm}
\usepackage{mathrsfs}
\usepackage{epsfig,psfrag}
\usepackage{amsmath,amsfonts,amssymb}

\DeclareMathOperator{\ch}{ch}
\DeclareMathOperator{\sh}{sh}

\newcommand{\Bcal}{\mathcal B}

\newcommand{\Ecal}{\mathcal E}

\newcommand{\Gcal}{\mathcal G}

\newcommand{\Mcal}{\mathcal M}

\newcommand{\Ocal}{\mathcal O}

\newcommand{\Zcal}{\mathcal Z}

\newcommand{\Dsc}{\mathscr D}

\newcommand{\nbf}{\mathbf {n}}

\newcommand{\pbf}{\mathbf {p}}

\newcommand{\qbf}{\mathbf {q}}

\newcommand{\rbf}{\mathbf {r}}

\newcommand{\xbf}{\mathbf {x}}

\newcommand{\ybf}{\mathbf {y}}

\newcommand{\ra}{\rightarrow}

\newcommand{\diag}{\mathrm{diag}}

\newcommand{\abs}[1]{| #1 |}

\newcommand{\str}[1]{\stackrel{#1}}

\newcommand{\til}[1]{\tilde{#1}}

\newcommand{\pd}{\partial}

\begin{document}

\preprint{draft}

\title{Non--perturbative approach to Casimir interactions in 
periodic geometries}

\author{Rauno B\"uscher and Thorsten Emig}

\affiliation{Institut f\"ur Theoretische Physik, Universit\"at zu
K\"oln, Z\"ulpicher Stra\ss e 77, 50937 K\"oln, Germany}

\date{\today}

\begin{abstract}
  Due to their collective nature Casimir forces can strongly depend on
  the geometrical shape of the interacting objects. We study the
  effect of {\it strong} periodic shape deformations of two ideal
  metal plates on their quantum interaction. A non-perturbative
  approach which is based on a path integral quantization of the
  electromagnetic field is presented in detail. Using this approach,
  we compute the force for the specific case of a flat plate and a
  plate with a rectangular corrugation. We obtain complementary
  analytical and numerical results which allow us to identify two
  different scaling regimes for the force as a function of the mean
  plate distance, corrugation amplitude and wave length. Qualitative
  distinctions between transversal electric and magnetic modes are
  revealed.  Our results demonstrate the importance of a careful
  consideration of the non-additivity of Casimir forces, especially in
  strongly non-planar geometries.  Non-perturbative effects due to
  surface edges are found. Strong deviations from the commonly used
  proximity force approximation emerge over a wide range of
  corrugation wave lengths, even though the surface is composed only
  of flat segments.  We compare our results to that of a perturbative
  approach and a classical optics approximation.
\end{abstract}

\pacs{03.70.+k, 11.10.-z, 42.50.Ct, 12.20.-m}

\maketitle

\section{Introduction}

Casimir interactions
\cite{H.B.G.Casimir,milonni,Bordag-Mostepanenko,Milton01,Kardar-Golestanian}
are a fundamental property of the vacuum.  They are commonly related
to quantum electrodynamics but fluctuation induced interactions are of
interest in a plethora of other fields like in condensed matter
systems as liquid crystals and superfluids
\cite{bwIsraelachvili,Garcia-Chan}, in cosmological models
\cite{Bytsenko-Cognola-et-al}, in particle physics
\cite{Milton80a,Milton80b} and in biological systems as proteins on
membranes. The guiding mechanism behind all these phenomena is that a
quantum or thermal field is constraint by boundary conditions on
surfaces so that the energy is modified, and effective interactions
between the surfaces occur. For quantum fields the Casimir interaction
is given by the {\it change} in the ground state energy $ {\cal
  E}_0=\frac{1}{2} \sum_n \hbar \omega_n$ due to the presence of
boundary conditions.  Even for non-interacting fields like the photon
gauge field it is difficult to obtain the Casimir interaction since
the eigenfrequencies $\omega_n$ can depend strongly on the confining
geometry. Thus it is not unexpected that exact analytical results for
Casimir interactions between macroscopic objects are not known even if
the geometry has high symmetry.

Most of the recent high precision experiments aim at the measurement
of the Casimir force in geometries which are closely related to the
standard case of two parallel plates
\cite{Lamoreaux,Mohideen-Roy,Chan-et-al,Bressi-et-al}. To the latter
case applies Casimir's seminal prediction \cite{H.B.G.Casimir}
\begin{equation}
\label{eq:original-force}
\frac{F_\text{flat}}{A}=-\frac{\pi^2}{240} \frac{\hbar c}{H^4}
\end{equation}
for the force between two ideal metallic and parallel plane plates of
area $A$ and distance $H$ at zero temperature.  For technical reasons,
usually a plate--sphere geometry is used in experiments. Even for this
case only approximative methods like the proximity force
or Derjaguin approximation \cite{Derjaguin} can be applied and the
exact result is not known. There exists geometries for which there is
even little intuition as to whether the interaction is attractive or
repulsive. A striking example is Boyer's result that the Casimir
energy of a conducting sphere is {\it positive} \cite{Boyer}. This
observation has triggered a search for repulsive configurations
\cite{Iannuzzi+03,Maclay00}. Such effects can be even of direct
practical relevance in nano technology where ``sticking'' of mobile
components in micromachines might be caused by Casimir forces
\cite{Serry-Walliser}.

The advances in experimental techniques have stimulated the
measurement of the shape dependence of Casimir forces in specially
designed geometries (as opposed to inevitable geometrical effects such as
surface roughness). Mohideen {\it et al.} were able to measure the
Casimir force between a sphere of large curvature radius and a
corrugated plate \cite{Roy-Mohideen,Chen-Mohideen-Klim-Mostepenner}.
Although the corrugation length was larger than the studied range of
separations between the surfaces, their results showed a clear
deviation from predictions of the proximity force approximation. While
it has been suggested \cite{Klim-Zanette-Caride} that lateral surface
displacements caused this discrepancy, there is no reason to believe
in the validity of the proximity approximation if the corrugation
length is decreased. 

Because of the wide range of realizations of Casimir forces, improved
experimental techniques and the increasing importance of
nanostructures it is interesting to develop novel approaches for
computing such interactions. In the limit of slight surface
deformations a path integral quantization subject to boundary
conditions allows for a perturbative calculation of the interaction
\cite{golkar-letter,golkar}, showing strong corrections to the
proximity approximation
\cite{Emig-Hanke-Golestanian-Kardar,Emig-Hanke-Golestanian-Kardar-II}.
Another perturbative approach, based on a multiple scattering
expansion, has been applied to the limit of large surface separations
\cite{Balian-Duplantier}.  Very recently, an alternative approximation
scheme based on geometric optics has been proposed for geometries
where the Casimir interaction is mostly caused by short wavelengths
\cite{jaffe}. However, to date no systematic method is known for
estimating interactions of strongly deformed objects, including large
curvature or even sharp edges.  In this paper we present a novel
non-perturbative method to compute electrodynamic Casimir interactions
between uniaxially and periodically deformed surfaces. It is based on
a path integral approach for Casimir forces \cite{likar-letter,likar}.
The approach is not restricted to small deformations or small surface
curvature but it also allows us to study strong deformations and
edges. We develop a numerical implementation of the approach which
allows for a precise computation of the interaction without any
approximations. As an example we consider a geometry consisting of a
flat and a rectangular corrugated plate, see Fig.\ref{eq:fig1}. For
this geometry we obtain the Casimir force over a wide range of surface
separations and corrugation lengths. We find that the edges of the
corrugated surface cause strong deviations from the proximity
approximation which agrees reasonably with our results only if the
corrugation length is much larger than both the surface distance and
the corrugation amplitude.  We show that the qualitative effect of
edges on the interaction can be understood in the limit of large
corrugation lengths in terms of classical ray optics.  A brief account
of our method and its application to scalar fields subject to
Dirichlet boundary conditions appeared in Ref.~\onlinecite{emig}.

The rest of this work is organized as follows. In Sec.~II we review
briefly the path integral approach and then introduce the method for a
non-perturbative computation of Casimir interactions. We consider
periodic uniaxially deformed surfaces. The new approach is then
applied in Sec.~III to the example of a flat and a corrugated surface
with sharp edges. The asymptotic limits of small and large corrugation
lengths are treated analytically. For arbitrary corrugation lengths
the interaction is obtain by an numerical implementation of our
approach. We give detailed numerical results for the total
electromagnetic Casimir force and the contributions from TM and TE
modes separately. In Sec.~IV we compare our results to perturbation
theory for slightly deformed smooth surfaces. We interpret our results
for large corrugation length in terms of classical ray optics.
Throughout the paper we set $c=1$ and $\hbar=1$.


\section{Non--perturbative path integral approach}
\label{sec:Non-pert-pi}

We consider two perfectly conducting periodically deformed
(corrugated) plates $S_\alpha$, ($\alpha=1,2$) with a mean separation
$H$.  They are assumed to be infinitely extended over a base plane
which is parameterized by the coordinates $\xbf_\|=(x_1,x_2)$. For
simplicity we assume that the corrugation is uniaxial along the $x_1$
direction. The shape of the plates is then described by height
functions $h_\alpha(x_1)$ which measure deviations from the mean
height so that $\int_{x_1}h_\alpha(x_1)=0$. The Casimir energy of the
two plate configuration can be obtained from an imaginary time path
integral representation \cite{golkar-letter,golkar} for the partition
function of the electromagnetic field and the confining plates. In the
absence of boundaries, the path integral extends over the
electromagnetic gauge field $A_\mu$ with the 4D space time action
\begin{equation}\label{eq:em-action}
S_0[A^\mu]\:=\:\frac{1}{4}\int d^4 X\,F_{\mu\nu}F^{\mu\nu}.
\end{equation}
and the field $F_{\mu\nu}=\pd_\mu A_\nu-\pd_\nu A_\mu$ and
$X=(x_0,\xbf_\|,x_3)$. In order to eliminate redundant gauge field
configurations the Fadeev Popov gauge fixing procedure has to be
applied \cite{peskin}.  The ideal metal boundary condition for the
gauge field $A_\mu(X)$ is given by the requirement that the tangential
components of the electric field vanishes at the surfaces.

For plate deformations which are uniaxial, the translational invariant
direction can serve as reference axis for defining transverse magnetic
(TM) and transverse electric (TE) modes, similar to the treatment of
wave guide geometries \cite{jackson}. Then every field configuration
can be decomposed into those two types of modes, and one can resort to
a scalar field path integral quantization
\cite{Emig-Hanke-Golestanian-Kardar,Emig-Hanke-Golestanian-Kardar-II}.
The scalar fields are given by the electric and magnetic field
component along the translational symmetry axis,
\begin{eqnarray}\label{eq:scalarfield}
\Phi(X) &=& E_2(X)\quad \mbox{for TM modes},\\
\Phi(X) &=& B_2(X)\quad \mbox{for TE modes}.
\end{eqnarray}

Since the plates are assumed to be ideally conducting, the boundary
conditions for TM and TE modes are of Dirichlet and Neumann type,
respectively, i.e.,
\begin{eqnarray}\label{eq:boundary-condition}
\Phi\big\vert_{S_\alpha} &=& 0 \quad \mbox{for TM modes},\\
\pd_{\hat{\nbf}_\alpha}\!\Phi\big\vert_{S_\alpha} &=& 0
\quad \mbox{for TE modes},
\end{eqnarray}
with the surface normal derivative denoted by $\pd_{\hat{\nbf}}$
pointing into the vacuum between the plates.  After a Wick rotation to
imaginary time $x_0\to ix_0$, both types of modes are described by the
Euclidean action
\begin{equation}\label{eq:e-action}
S_E\{\Phi\} = \frac{1}{2}\int d^4 X \, (\nabla \Phi)^2.
\end{equation}
In 4D Euclidean space, the surface positions of the plates are then
parameterized by
$X_\alpha(\rbf)=\left[\rbf,h_\alpha(x_1)+H\delta_{\alpha2}\right]$
with $\rbf\equiv (x_0,\xbf_\|)$.  Following the procedure introduced in
Refs.~\onlinecite{likar-letter,likar}, the boundary conditions are
imposed by inserting delta functions on the surface in the functional
integral.  The partition function for TM and TE modes,
respectively, then reads
\begin{eqnarray}
\Zcal_\text{D}&=&\Zcal_0^{-1}\int\Dsc\Phi\prod_{\alpha=1}^2 \prod_{X_\alpha}
\delta\left[\Phi(X_\alpha)\right]\,
e^{-S_E\{\Phi\}},\\
\Zcal_\text{N}&=&\Zcal_0^{-1}\int\Dsc\Phi\prod_{\alpha=1}^2 \prod_{X_\alpha}
\delta\left[\pd_{\hat{\nbf}_\alpha}\!\Phi(X_\alpha)\right]\,
e^{-S_E\{\Phi\}}
\end{eqnarray}
with the boundary free partition function $\Zcal_0$.  The functional
integrals can be calculated by introducing auxiliary fields to
represent the delta functions.  Then, the Gaussian integration over
$\Phi$ can be carried out, yielding the partition function in terms of
an effective action for the auxiliary fields, 
\begin{equation}\label{eq:effective-action}
\Zcal=\int\prod_{\alpha=1}^2\Dsc \psi_\alpha \,
e^{-S_\text{eff}\{\psi_\alpha\}}
\end{equation}
with the effective action 
\begin{equation}\label{eq:effective-action-2}
S_\text{eff}\{\psi_\alpha\}=\frac{1}{2}
\int_\rbf \!\int_{\rbf'}\! \sum_{\alpha,\beta}
\psi_\alpha(\rbf)\Mcal^{\alpha\beta}(\rbf;\rbf')
\psi_\beta(\rbf').
\end{equation}
In the following we will drop the subscript D or N for the boundary
conditions on all quantities which apply to both conditions in the
same way.  The total electrodynamic Casimir energy is then given by
the sum of TM and TE mode contributions,
$\Ecal=\Ecal_\text{TM}+\Ecal_\text{TE}$.  After subtracting the
divergent and $H$ independent terms, the energies can be written as
$\Ecal_\text{TM}=\ln\det(\Mcal_\text{D}\Mcal^{-1}_{\text{D},\infty})/(2AL)$,
and analogous for $\Ecal_\text{TE}$ with $D$ replaced by $N$, where
$\Mcal_{\text{D},\infty}$ is the asymptotic expression of
$\Mcal_\text{D}$ for $H\to\infty$, $A$ is the surface area of the
plates, and $L$ is the Euclidean length in time direction.  The
Casimir force $F=-\pd_H \Ecal$ per unit area is then given by
$F=F_\text{TM}+F_\text{TE}$ with
\begin{subequations}
\label{eq:force}
\begin{eqnarray}
F_\text{TM}&=&-\frac{1}{2AL}\,\text{Tr}\left(\Mcal_\text{D}^{-1}
\pd_H\Mcal_\text{D}\right),\\
F_\text{TE}&=&-\frac{1}{2AL}\,\text{Tr}\left(\Mcal_\text{N}^{-1}
\pd_H\Mcal_\text{N}\right).
\end{eqnarray}
\end{subequations}
The right hand side of these expressions is always finite, and no
regulation of divergences by subtraction of the vacuum energy in the
absence of boundaries is necessary.  The Dirichlet and Neumann
matrix kernels of the effective Gaussian action can be expressed in
terms of the Euclidean scalar Green function
$\Gcal(\rbf,x_3)=(\rbf^2+x_3^2)^{-1}/(4\pi^2)$, and are respectively
given by
\begin{subequations}
\label{eq:matrix-kernels}
\begin{eqnarray}
\label{eq:Dirichlet-matrix-kernel}
\Mcal_{\text{D}}^{\alpha\beta}\left(\rbf;\rbf'\right) &=& 
\eta_\alpha(x_1)\,\eta_\beta(x'_1)\,
\Gcal\left(X_\alpha(\rbf)-X_\beta(\rbf')\right),\\
\label{eq:Neumann-matrix-kernel}
\Mcal_{\text{N}}^{\alpha\beta}\left(\rbf;\rbf' \right) &=& 
\eta_\alpha(x_1)\,\eta_\beta(x'_1)\,
\pd_{\hat{\nbf}_\alpha(x_1\!)}\pd_{\hat{\nbf}_\beta(x'_1\!)}
\Gcal\left(X_\alpha(\rbf)-X_\beta(\rbf')\right)
\end{eqnarray}
\end{subequations}
with the coefficients given by $\eta_\alpha(x_1)=
\left(1+[(\pd_{x_1}\!h_\alpha)(x_1)]^2\right)^{\frac{1}{4}}$. These
coefficient arise from the integral measure on the curved surfaces.
However, since they are independent of the mean plate distance $H$,
they cancel in the matrix product of Eq. (\ref{eq:force}) and
therefore can be ignored for the calculation of forces. The kernels
are symmetric with
$\Mcal(\rbf;\rbf')=\Mcal^\text{T}(\rbf';\rbf)$
where the transpose refers to $\alpha$, $\beta$.  Using the
parameterization in terms of height profiles, the matrix kernels can
now be written as
\begin{subequations}
\label{eq:matrix-M}
\begin{eqnarray}\label{eq:Dirichlet-matrix-M}
\Mcal^{\alpha\beta}_\text{D}\left(\rbf;\rbf'\right)&=& 
\Gcal\left(\rbf-\rbf',h_\alpha(x_1)-h_\beta(x'_1)
+H(\delta_{\alpha2}-\delta_{\beta2})\right),\\
\label{eq:Neumann-matrix-M}
\Mcal^{\alpha\beta}_\text{N}\left(\rbf;\rbf'\right)&=&
(-1)^{\alpha+\beta}
\left[-\pd_{x_3}^2+\left(h_\alpha'(x_1)+h_\beta'(x'_1)\right)\,
\pd_{x_1}\pd_{x_3}-h_\alpha'(x_1)h_\beta'(x'_1)\,\pd_{x_1}^2\right]
\nonumber\\
&&\quad\times\:
\Gcal\left(\rbf-\rbf',x_3-x'_3\right)
\Big\vert_{\begin{smallmatrix}
x_3=h_\alpha(x_1)+H\delta_{\alpha2}\\x'_3=h_\beta(x'_1)+H\delta_{\beta2}
\end{smallmatrix}}
\end{eqnarray}
\end{subequations}
for the Dirichlet and Neumann case, respectively. So far, we have not
used the periodicity of the surface profile, and the above results are
valid for any uniaxial deformation. However, the computation of the
force can be performed more efficiently if the periodic symmetry of
the surface is used. 

\begin{figure}[t]
\includegraphics[width=0.99\linewidth]{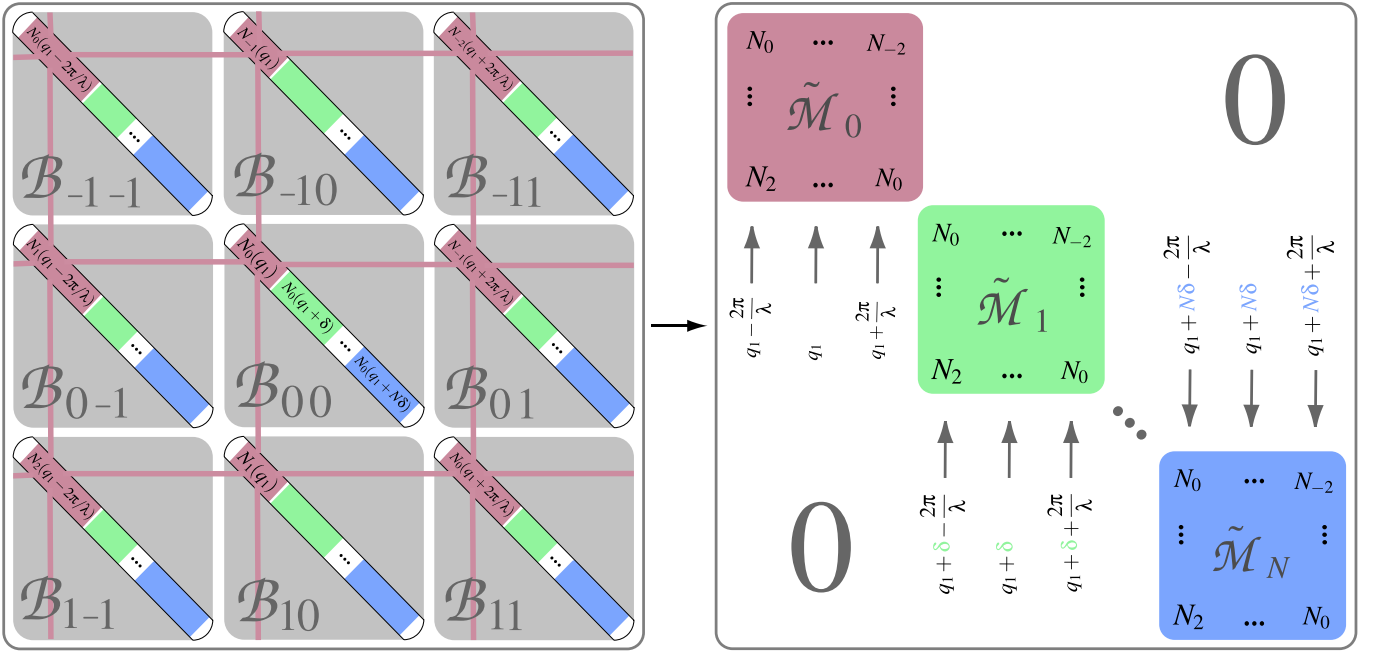}
\caption{\label{eq:trafo} (color online)
  Transformation of the matrix $\til{\Mcal}$ to
  block-diagonal form. The figure shows a finite part of the matrix,
  corresponding to the blocks $\Bcal_{kl}$ with $k$, $l=-1,0,1$,
  before and after the permutations of rows and columns. Before the
  transformation (left box) $\til{\Mcal}$ has a band
  structure with diagonal blocks $\Bcal_{ij}$ consisting of $2 \times
  2$ matrices $N_m$ along the diagonal. (The dependence on the lateral
  momentum $\qbf_\perp$ is not shown here.) The first step of the
  transformation is to permute the rows and columns which are formed
  by the first entry $N_m$ in every block $\Bcal_{kl}$ (indicated as
  grid).  These entries from after the permutations the first block
  $\til \Mcal_0$ of $\til{\Mcal}$ (right box). The latter
  permutation process is then repeated for the second entry, the third
  entry until the $(N+1)^\text{th}$ entry of every block $\Bcal_{kl}$,
  leading to the $N+1$ blocks $\til \Mcal_j$. The momenta $q_1$ within
  each block $\til \Mcal_j$ are constant for every column and they differ
  only by integer multiples of $2\pi/\lambda$ between columns (of the
  same block), see labels in the right box. The blocks $\til \Mcal_j$
  differ in their momentum shift $j\delta$, $\delta=2\pi/W$, which is
  located in the unit cell $[0,2\pi/\lambda[$ since $j=0, \ldots ,
  N=W/\lambda-1$.}
\end{figure}

Due to the translational invariance in time ($x_0$) and one space
($x_2$) direction, it is convenient to introduce the momentum vector
$\qbf_\perp=(q_0,q_2)$ which is perpendicular to the direction of
modulation.  Due to the periodicity of the surface profile, the
Fourier transform $\til{\Mcal}(\pbf;\qbf)=\int_\rbf
\int_{\rbf'} e^{i\pbf\rbf+i\qbf\rbf'} \Mcal(\rbf;\rbf')$ can
be decomposed into the series
\begin{equation}\label{eq:Bloch}
\til{\Mcal}\left(p_1,\pbf_\perp;q_1,\qbf_\perp\right)\:=\:
(2\pi)^3\delta\left(\pbf_\perp\!+\qbf_\perp\right)\,
\sum_{m=-\infty}^\infty
\delta\left(p_1+q_1+2\pi m/\lambda\right)\,
N_m\left(q_\perp,q_1\right)
\end{equation}
where $N_m\left(q_\perp,q_1\right)$ are $2\times2$ matrices which
depend only on $q_\perp=\abs{\qbf_\perp}$. From Eq.~(\ref{eq:Bloch})
it is obvious that the matrix $\til{\Mcal}$ has its non-zero entries
arranged in $2 \times 2$ blocks along parallel bands. Due to this
structure, there exists a transformation, consisting only of row and
column permutations, which makes the matrix block diagonal. To perform
this transformation, we cut the matrix $\til{\Mcal}$ into smaller
matrices $\Bcal_{kl}$ which have non-zero entries only in $2 \times 2$
blocks along the diagonal, see Fig.\ref{eq:trafo}. For the purpose of
parameterization, we consider discrete momenta $p_1=(2\pi/W) j$,
$j=0,\ldots,N$, along the direction of surface modulation with
$N=W/\lambda-1$. The continuum limit is obtained if the linear size
$W$ of the surfaces and $N$ are taken to infinity in order to obtain
the force per surface area $A=W^2$. With this parameterization, the
block diagonal matrices $\Bcal_{kl}$ of dimension $2(N+1)\times
2(N+1)$ can be read off from Eq.~(\ref{eq:Bloch}), leading to
\begin{equation}\label{eq:Bloch-4}
\Bcal_{kl}(q_\perp)\:=\:
\diag\{ B_{kl}\left(q_\perp, 0 \right), B_{kl}\left(q_\perp,2\pi/W \right), 
\dots , B_{kl}\left(q_\perp,2\pi N/W \right) \}
\end{equation}
with the $2 \times 2$ block matrices defined as [see Fig.~\ref{eq:trafo}]
\begin{equation}
\label{eq:matrix-B}
B_{kl}\left(q_\perp,q_1\right)\:=\:
N_{k-l}\left(q_\perp,q_1+2\pi l/\lambda\right).
\end{equation}

By inspection of Fig.~\ref{eq:trafo} one easily realizes that a
sequence of row permutations and a subsequent sequence of column
permutations transforms the matrix $\til{\Mcal}$ to
block-diagonal form. Each of the $N+1$ blocks $\til{\Mcal}_j$ is
composed of exactly one element from each matrix $\Bcal_{kl}$ and
those elements forming a block $\til{\Mcal}_j$ come from the same
position in every matrix $\Bcal_{kl}$ as indicated by the color scheme
in Fig.~\ref{eq:trafo}. Thus, each block $\til{\Mcal}_j$ is composed
of entries which correspond to the same discrete momentum
$p_1=(2\pi/W) j$, and we obtain for the elements of $\til{\Mcal}_j$
the result
\begin{equation}
\label{eq:block-elements}
\til{\Mcal}_{j,kl}\left(\pbf_\perp,\qbf_\perp\right)
\:=\:(2\pi)^2\,\delta\left(\pbf_\perp\!+\qbf_\perp\right) 
B_{kl}\left(q_\perp,2\pi j/W\right).
\end{equation}
The number of permutations needed for the matrix transformation is
always even, and thus we get the determinant
\begin{equation}
\label{eq:block-decomp}
\det\til{\Mcal}\:=\:\prod_{j=0}^N\det\til{\Mcal}_j.
\end{equation}
By differentiating with respect to the mean surface distance $H$ and
by using the relation $\ln \det \til{\Mcal}_j = \text{Tr} \ln
\til{\Mcal}_j$, we obtain
\begin{equation}
\label{eq:sum-subsystems}
\pd_H \left(\ln\det\til{\Mcal}\right)\:=\:\sum_{j=0}^N{\text{Tr}} 
\left(\til{\Mcal}_j^{-1}\cdot\pd_H\til{\Mcal}_j\right)\,.
\end{equation}
This result reflects the fact that the (free) energy of the system can
be calculated as the sum of the individual (free) energies of {\it
  decoupled} subsystems, which are described by the matrices
$\til{\Mcal}_j$. Each subsystem with fixed $j$ describes scattering
events at the {\it fixed} momenta $p_1=(2\pi/W)j + (2\pi/\lambda) l$
which differ only by integer multiples of $2\pi/\lambda$.

Using Eq.~(\ref{eq:block-elements}) we can perform the trace over the
continuous lateral momenta and the discrete indices within a fixed
subsystem,
\begin{equation}
\label{eq:trace}
\text{Tr} \left( \til{\Mcal}_j^{-1}\cdot\pd_H\til{\Mcal}_j \right) = 
\frac{LW}{(2\pi)^2} \int d^2 \qbf_\perp  
\sum_{\begin{smallmatrix}
k,l=-\infty\\
\alpha,\beta=1,2
\end{smallmatrix}}^\infty
B^{-1}_{kl,\alpha\beta}\left(q_\perp,q_1\right)
\cdot \pd_H B_{lk,\beta\alpha}\left(q_\perp,q_1\right).
\end{equation}
where we have explicitly indicated that the trace is performed with
respect to all discrete indices, and we remind that $L$ is the system
size in time direction. It appears useful to define the function
\begin{equation}\label{eq:g-fct}
g\left(q_\perp,q_1\right)\equiv\text{tr}\left(B^{-1}\!\left(q_\perp,q_1\right) 
\cdot\pd_H B\left(q_\perp,q_1\right)\right),
\end{equation}
with the lower-case symbol tr denoting the trace over the discrete
indices summed over in Eq.~(\ref{eq:trace}). Next we perform the sum
over all subsystems with $j=0, \ldots, N=W/\lambda-1$. This can be
easily done by going back to continuous momenta $p_1$. If we take the
limit $W$, $N \to \infty$ with $W/(N+1)=\lambda$ fixed the sum in
Eq.~(\ref{eq:sum-subsystems}) can be written as the integral
\begin{equation}
\pd_H\left(\ln\det \Mcal\right)\:=\:\frac{LW}{(2\pi)^2}\int
d^2 \qbf_\perp \,\frac{W}{2\pi}\, \int_0^{2\pi/\lambda} dq_1 \, 
g\left(q_\perp,q_1\right).
\end{equation}
The function $g(q_\perp,q_1)$ has the following symmetry properties. A
shift of the momentum $p_1$ by $2\pi/\lambda$ corresponds just to a
renumbering of the matrix elements $B_{kl}$ since the matrix is of
infinite dimension. Thus we have
$g(q_\perp,q_1+2\pi/\lambda)=g(q_\perp,q_1)$. If both surface profiles
are described by even functions, $h_\alpha(-x_1)=h_\alpha(x_1)$, for
the matrices $N_m$ the relation
$N_m(q_\perp,-q_1)=N_{-m}(q_\perp,q_1)$ holds. Using the later
relation and the definition of $B_{kl}$ of Eq.~(\ref{eq:matrix-B}) it
is easy to check that $g(q_\perp,-q_1)=g(q_\perp,q_1)$ by performing
appropriate row and column permutations for the matrix $B$. The above
symmetries allow to write the Casimir force per unit area,
$F/A=-(1/2LW^2)\,\pd_H(\ln \det\Mcal)$, as
\begin{equation}\label{eq:force-result}
F/A\:=\:-\frac{1}{4\pi^2}\int_0^\infty
dq_\perp q_\perp\int_0^{\pi/\lambda} dq_1\,g\left(q_\perp,q_1\right).
\end{equation}
This is the final result of the general approach for arbitrary
uniaxially corrugated surfaces. As we will show below, it can be
used for an efficient numerical computation of the Casimir force.  The
input of such a numerical approach are the matrices $N_m$ from the
decomposition in Eq.(\ref{eq:Bloch}).  Moreover, the result can be
also used to obtain non--perturbative analytical results in the
asymptotic limit of very small corrugation lengths.

Before one can develop a numerical implementation of the above
representation of the Casimir force, one of course has to restrict the
infinite dimensional matrices. In the remaining part of this section
we will introduce a suitable cutoff procedure for the matrix
dimension.  We will take two flat plates as a simple example to
examine the convergence of the procedure if the cutoff is taken to
large values.  The cutoff procedure consists in the restriction of the
matrix $\til \Mcal$ to blocks $\Bcal_{kl}$ with $k$,
$l=-M,\ldots,M$ only. The dimension of $\til \Mcal$ is then
$2(2M+1)(N+1)$.  Fig.~\ref{eq:trafo} displays the restricted matrix
for $M=1$.  The corresponding function $g$ is then defined by
Eq.~(\ref{eq:g-fct}) with the restriction that the trace runs over
$k$, $l=-M,\ldots,M$ only.  We will denote this function in the
following by $g_M$. This function is then used instead of $g$ in
Eq.~(\ref{eq:force-result}) to obtain a series of approximations $F_M$
to the force which converges to $F$ for $M \to \infty$. As an example
consider two flat plates at distance $H$. Then $\til \Mcal$
is a diagonal matrix and $N_m=0$ for $m \neq 0$. Thus the matrix $B$
is also diagonal with
\begin{equation}
  \label{eq:B-flat}
  B_{kl}(q_\perp,q_1)=\delta_{kl} \, N_0(q_\perp,q_1+2\pi l/\lambda).
\end{equation}
Using Eq.~(\ref{eq:g-fct}) with the trace taken for $k$, $l=-M,
\ldots, M$, one gets the function
\begin{equation}
  \label{eq:g-finite-flat}
  g_M(q_\perp,q_1)=\sum_{l=-M}^M \frac{2\sqrt{q_\perp^2+(q_1+2\pi l/\lambda)^2}}
{e^{2\sqrt{q_\perp^2+(q_1+2\pi l/\lambda)^2}H}-1}.
\end{equation}
Integration over $q_1$ yields an $M^\text{th}$ order approximation
$F_M$ to the force,
\begin{eqnarray}
  \label{eq:force-finite-flat}
  F_M/A &=& -\frac{1}{8\pi^2} \int_0^\infty dq_\perp q_\perp \int_0^{2\pi/\lambda}
 dq_1 g_M(q_\perp,q_1)\nonumber\\
& = & -\frac{1}{8\pi^2} \int_0^\infty dq_\perp q_\perp 
\int_{-2\pi M/\lambda}^{2\pi(M+1)/\lambda} dq_1 \frac{2q}{e^{2qH}-1}
\end{eqnarray}
with $q=\sqrt{q_\perp^2+q_1^2}$. For $M \to \infty$ one gets the known
($\lambda$ independent) result $F/A=-(\pi^2/480) H^{-4}$, and the
finite $M$ corrections to this asymptotic result scale exponentially
fast ($\sim e^{-4\pi M H/\lambda}$) to zero for large $M$.  Therefore,
in the case of periodically deformed plates one can expect accurate
numerical results for $F$ from moderate values for the cutoff $M$, and
the convergence is faster for smaller $\lambda$.


\section{Rectangular corrugation}

\begin{figure}[t]
\includegraphics[width=0.5\linewidth]{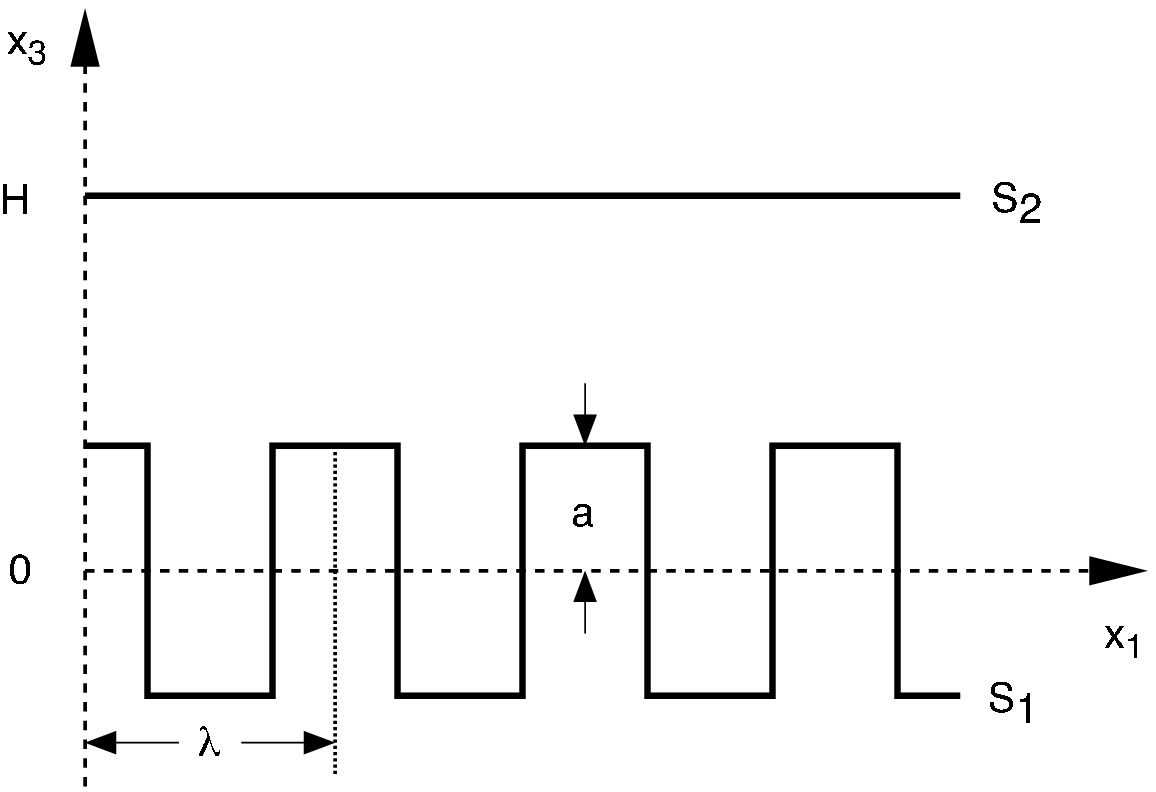}
\caption{\label{eq:fig1} Geometry consisting of a rectangular corrugated 
plate and a flat plate. The surfaces are translationally invariant along the
$x_2$ direction.}
\end{figure}

In the previous section we developed a non-perturbative approach for
computing Casimir interactions between periodically deformed surfaces.
In this section we will use the approach to obtain explicit results
for the Casimir force between a flat plate and a plate with a
rectangular grating. The effect of this class of periodic geometries
(corrugated surfaces) can significantly modify the interaction of the
objects \cite{Karepanov+87,Klimchitskaya+96}.  Is was proposed that
such geometries can be used to reveal novel features of the Casimir
interaction \cite{golkar-letter,golkar}. For a similar geometry
consisting of a sinusoidally corrugated plate and a sphere with a
radius $\gg H$ Roy and Mohideen measured the force, and found clear
deviations from the predictions of the proximity force approximation
\cite{Roy-Mohideen}.  While it has been suggested that lateral shifts
of the surfaces caused the discrepancy, we demonstrate below that
periodic surfaces allow for a much stronger sensitivity to geometry if
the corrugation length is reduced to smaller values. Specifically, we
consider the geometry shown in Fig.~\ref{eq:fig1} with a rectangular
grating of amplitude $a$ and wavelength $\lambda$.  Choosing $x_1$ as
the direction of modulation, this corresponds to the height profile
\begin{equation}\label{eq:profile}
h_1(x_1)\:=\:\left\{
\begin{array}{lll}
+a & \text{for} & |x_1| < \lambda/4\\
-a & \text{for} & \lambda/4 < |x_1| < \lambda/2
\end{array}\right.,
\end{equation}
and continuation by periodicity $h_1(x_1)=h_1(x_1+n\lambda)$ for any
integer $n$.  The upper plate is flat so that $h_2(x_1)=0$.

The main purpose of our work is to obtain the Casimir interaction in
regimes where other methods like proximity approximation, pairwise
summation of two body forces or perturbation theory fail or become
unreliable. While the proximity approximation assumes smooth profiles
with small local curvature also perturbation theory in the height
profile yields divergences in the presence of edges in the profile or
if the corrugation length becomes very small, i.e., $\lambda \ll a$,
$H$
\cite{Emig-Hanke-Golestanian-Kardar,Emig-Hanke-Golestanian-Kardar-II}.
In perturbation theory, finite corrections of order $a^2$ are
recovered only if edges are ``smeared out'' over a finite length
scale.  Thus the correct procedure would be presumably to sum all
orders of perturbation theory for a ``smeared out'' profile, and then
to take the limit of sharp edges {\it after} summing all
contributions. Since the perturbative treatment is rather cumbersome
the latter program is not practicable, and non-perturbative techniques
are imperative. 

In order to apply the general result of Eq.~(\ref{eq:force-result})
for the Casimir force, we have to decompose the matrix $\til{\Mcal}$
into the matrices $N_m$, see Eq.~(\ref{eq:Bloch}).  For a general
profile, this has to be done by a numerical Fourier transformation.  A
nice property of the rectangular profile of Eq.~(\ref{eq:profile}) is
that it allows for an analytical computation of the matrices $N_m$.
The idea is to rewrite the profile of the corrugated plate as a
discrete Fourier series,
\begin{equation}\label{eq:sfm-series}
h_1(x_1)\:=\:\frac{2a}{\pi}
\sum_{n=-\infty}^\infty \frac{(-1)^{n-1}}{2n-1}\,
e^{\frac{2\pi i}{\lambda}(2n-1)x_1}
\end{equation}
which is inserted into the matrix $\Mcal$ of Eq.~(\ref{eq:matrix-M}).
Then the Fourier transformed matrix $\til{\Mcal}$ can be calculated,
leading after some algebra to the matrices
$N_m\left(q_\perp,q_1\right)$.  Details of this calculation are given
in Appendix A.  The results are
\begin{equation}\label{eq:N_D-matrix}
N_{\text{D},m}\left(q_\perp,q_1\right)\:=\:\left\{
\begin{array}{ll}
\left(
\begin{array}{cc}
A^{\text{D}}_m\left(q_\perp,q_1\right) & 0\\[5pt] 0 & 0
\end{array}\right)\:+\:\delta_{m0}\left(\begin{array}{cc}
\frac{1}{4q}(1+e^{-2aq}) & \frac{e^{-qH}}{2q}\ch(aq)\\[5pt]
\frac{e^{-qH}}{2q}\ch(aq) & \frac{1}{2q}
\end{array}\right) & \mbox{for $m$ even}
\\[20pt]
\left(
\begin{array}{cc}
0 & \frac{(-1)^\frac{m-1}{2}}{m\pi}\frac{e^{-qH}}{q}\sh(aq)\\[5pt]
\frac{(-1)^\frac{m-1}{2}}{m\pi}
\frac{e^{-\til{q}_mH}}{\til{q}_m}\sh(a\til{q}_m) & 0
\end{array}\right) & \mbox{for $m$ odd}
\end{array}\right.
\end{equation}
for Dirichlet conditions and
\begin{equation}\label{eq:N_vN-matrix}
N_{\text{N},m}\left(q_\perp,q_1\right)\:=\:\left\{
\begin{array}{ll}
\left(
\begin{array}{cc}
A^{\text{N}}_m\left(q_\perp,q_1\right) & 0\\[5pt]
0 & 0
\end{array}\right)\:+\:\delta_{m0}\left(\begin{array}{cc}
-\frac{q}{4}(1+e^{-2aq}) & \frac{q}{2}e^{-qH}\ch(aq)\\[5pt]
\frac{q}{2}e^{-qH}\ch(aq) & -\frac{q}{2}
\end{array}\right) & \mbox{for $m$ even}
\\[20pt]
\left(
\begin{array}{cc}
0 & \frac{(-1)^\frac{m-1}{2}}{m\pi}\,e^{-qH}
\big[q+\frac{2\pi m}{\lambda}\frac{q_1}{q}\big]\sh(aq)\\[5pt]
\frac{(-1)^\frac{m-1}{2}}{m\pi}\,e^{-\til{q}_mH}
\big[\til{q}_m-\frac{2\pi m}{\lambda}
\frac{q_1+2\pi m/\lambda}{\til{q}_m}\big]
\sh(a\til{q}_m) & 0
\end{array}\right) & \mbox{for $m$ odd}
\end{array}\right.
\end{equation}
for Neumann conditions with
\begin{equation}
A^{\text{D}}_m\left(q_\perp,q_1\right)\:=\:\frac{1}{\pi^2}
\sum_{k=-\infty}^\infty
\frac{(-1)^\frac{m}{2}}{(2k-1)(m-2k+1)}\,
\frac{e^{-2a\til{q}_{2k-1}}-1}{\til{q}_{2k-1}}
\end{equation}
and
\begin{equation}
\begin{split}
A^{\text{N}}_m\left(q_\perp,q_1\right)\:&=\:\frac{1}{\pi^2}
\sum_{k=-\infty}^\infty
\frac{(-1)^\frac{m}{2}}{(2k-1)(m-2k+1)}\,
\frac{1-e^{-2a\til{q}_{2k-1}}}{\til{q}_{2k-1}^3}\\[5pt]
&\qquad\times\,
\left[q_1\big(q_1+\frac{2\pi m}{\lambda}\big)
\big(q_1+\frac{2\pi}{\lambda}(2k-1)\big)^2+2q_\perp^2
\big(q_1+\frac{\pi m}{\lambda}\big)\big(q_1+\frac{2\pi}{\lambda}
(2k-1)\big)+q_\perp^4\right],
\end{split}
\end{equation}
respectively, with the definition
$\til{q}_n=\sqrt{q_\perp^2+(q_1+2\pi n/\lambda)^2}$
which implies $q\equiv\til{q}_0$. With these results at hand, the
Casimir force can be calculated by the approach developed in the
previous section. The recipe is as follows. First, one constructs the
matrix $B_{kl}$ of Eq.~(\ref{eq:matrix-B}), then one calculates the
inverse of $B_{kl}$ to obtain the function $g(p_\perp,p_1)$ of
Eq.~(\ref{eq:g-fct}), and finally one has to perform the integration of
Eq.~(\ref{eq:force-result}).  In general, this program can only be
performed numerically.  However, in the limit $\lambda \to 0$ it turns
out that a closed form for the function $g(p_\perp,p)$ is available
which allows to obtain the Casimir force in this limit exactly.

\subsection{The limit of small $\lambda$}
\label{sec:small-lam}

Let us consider the case where the corrugation length $\lambda$ sets
the smallest length scale in the geometry of Fig.\ref{eq:fig1}. If we
take the extreme limit of $\lambda \to 0$, a naive assumption is that
the field can no longer get into the narrow valleys of the corrugated
plate. Even for small but finite $\lambda$ this picture should be a
good, though approximate, description since it still effects the
wavelengths of order $H$ which give the main contribution to the
force. Thus one expects that the plates feel a force which is equal to
the force between two {\it flat} plates at the {\it reduced} distance
$H-a$. However, the question remains to what extent this is a good
approximation when $\lambda$ becomes larger, say of order $a$.  To
check our naive expectation, we will apply the approach of the
previous section to the limit $\lambda \to 0$.  Fortunately, in this
limit the matrices $N_m(q_\perp,q_1)$ simplify considerably both for
TM and TE modes.  The explicit form of these matrices is given in
appendix B.  From this result, we can explicitly calculate the
functions $g_M(q_\perp,q_1)$ which was introduced before
Eq.~(\ref{eq:B-flat}).  As explained in section \ref{sec:Non-pert-pi},
the infinite dimensional matrix $B_{kl}$ is truncated for the
calculation at order $M$ with $k,l=-M,\ldots,M$ so that the truncation
is done symmetrically around the center at $(k,l)=(0,0)$ which
contains the leading matrix entries. From the exponential convergence
behavior of the flat plate result given below
Eq.~(\ref{eq:force-finite-flat}) one can expect that in the extreme
limit $\lambda \to 0$, the series $g_M(q_\perp,q_1)$ converges so
rapidly towards $g(q_\perp,q_1)$ that already for $M=1$ the exact
asymptotic expression is obtained. Indeed, our explicit calculation of
$g_M(q_\perp,q_1)$ for low $M$ confirms this expectation.  From the
truncated matrix $B_{kl}$ of Eq.~(\ref{eq:matrix-B}) and the matrices
of appendix B we get the simple result
\begin{equation}
g_M(q_\perp,q_1)\:=\:\left\{
\begin{array}{cc}
-\frac{2q(1+e^{-2aq})}{1+e^{-2aq}-2e^{2(H-a)q}} & 
\mbox{\text{for} $M=0$}\\[5pt]
q\left[\coth\left(q(H-a)\right)-1\right] & \mbox{\text{for} $M\ge 1$}
\end{array}\right.
\end{equation}
for both TM and TE modes.  Thus from first order ($M=1$) on the
function $g_M(q_\perp,q_1)$ {\it remains invariant} with increasing
dimension $M$ of the matrix $B_{kl}$. Interestingly, the result for
$M\ge 1$ has precisely the form, which one gets for two {\it flat }
plates at reduced distance $H-a$.  If one integrates the function
$g_M(q_\perp,q_1)$ for $M=1$ one obtains from
Eq.~(\ref{eq:force-result}) the Casimir force per surface area
\begin{equation}
\label{eq:F_0}
F_0/A \:=\: -\frac{\pi^2}{480}\frac{1}{(H-a)^4}
\end{equation}
for both TM and TE modes. Thus in the limit $\lambda \to 0$ both types
of modes yield the same contribution to the total electrodynamic
Casimir force $F=2F_0$. The result of Eq.~(\ref{eq:F_0}) corresponds to
the naive reduced distance argument given at the beginning of this
section.  Notice that this result is non--perturbative in $a/H$ and is
{\it exact} in the limit $\lambda \to 0$.  Perturbation theory for
smoothly deformed surfaces always yields corrections to the force of
order $a^2$
\cite{Emig-Hanke-Golestanian-Kardar,Emig-Hanke-Golestanian-Kardar-II}.
However, for small $a/H$, the result of Eq.~(\ref{eq:F_0}) has the
expansion
\begin{equation}
F_0/A \:=\: - \frac{\pi^2}{480}\frac{1}{H^4}
\left[1+4\frac{a}{H}+\Ocal\left(\left(\frac{a}{H}\right)^2\right)\right]
\end{equation}
which indicates that perturbation theory is not applicable if $\lambda
\ll a$.  Below we will see that the force $F_0$ provides an {\it upper
  bound} for the Casimir force from both TM and TE modes at fixed
$H/a$, i.e., for increasing $\lambda$ the force always decreases
compared to $F_0$. We expect that the results of this section for
$\lambda \to 0$ are valid for corrugations of arbitrary shape and also
for rough surfaces if $\lambda$ is identified with the characteristic
length scale for surface deformations.

\subsection{The limit of large $\lambda$} 

In the opposite limit of very large $\lambda$ the corrugated surface
is composed of large flat segments with a low density of edges. At
sufficiently small surface separations $H \ll \lambda$ the main
contribution to the force comes from wavelengths which are much
smaller than the scale $\lambda$ of the surface structure. Thus in the
dominant range of modes diffraction can be neglected, and the simple
proximity force approximation (Derjaguin approximation
\cite{Derjaguin}) should be applicable.  Such an approximation assumes
that the total force can be calculated as the sum of local forces
between opposite {\it flat} and {\it parallel} small surface elements
at their local distance $H-h(x_1)$. No distinction is made between TM
and TE modes.  This procedure is rather simple for the rectangular
corrugation considered here since the surface has no curvature (except
for edges). There are only two different distances $H+a$, $H-a$ which
contribute one half each across the entire surface area, leading for
$\lambda \to \infty$ to the proximity approximation for the force,
\begin{equation}
\label{eq:F_infty}
F_\infty/A\:=\:-\frac{\pi^2}{480}\,
\frac{1}{2}
\left[\frac{1}{(H-a)^4}+\frac{1}{(H+a)^4}\right].
\end{equation}
Below we will see that later result provides a {\it lower bound} for
the Casimir force from both TM and TE modes. In contrast to the limit
of small $\lambda$ the correction for small $a/H$ is of order
$(a/H)^2$ here.

\subsection{Numerical analysis}

\begin{figure}[h!]
\includegraphics[width=0.48\linewidth]{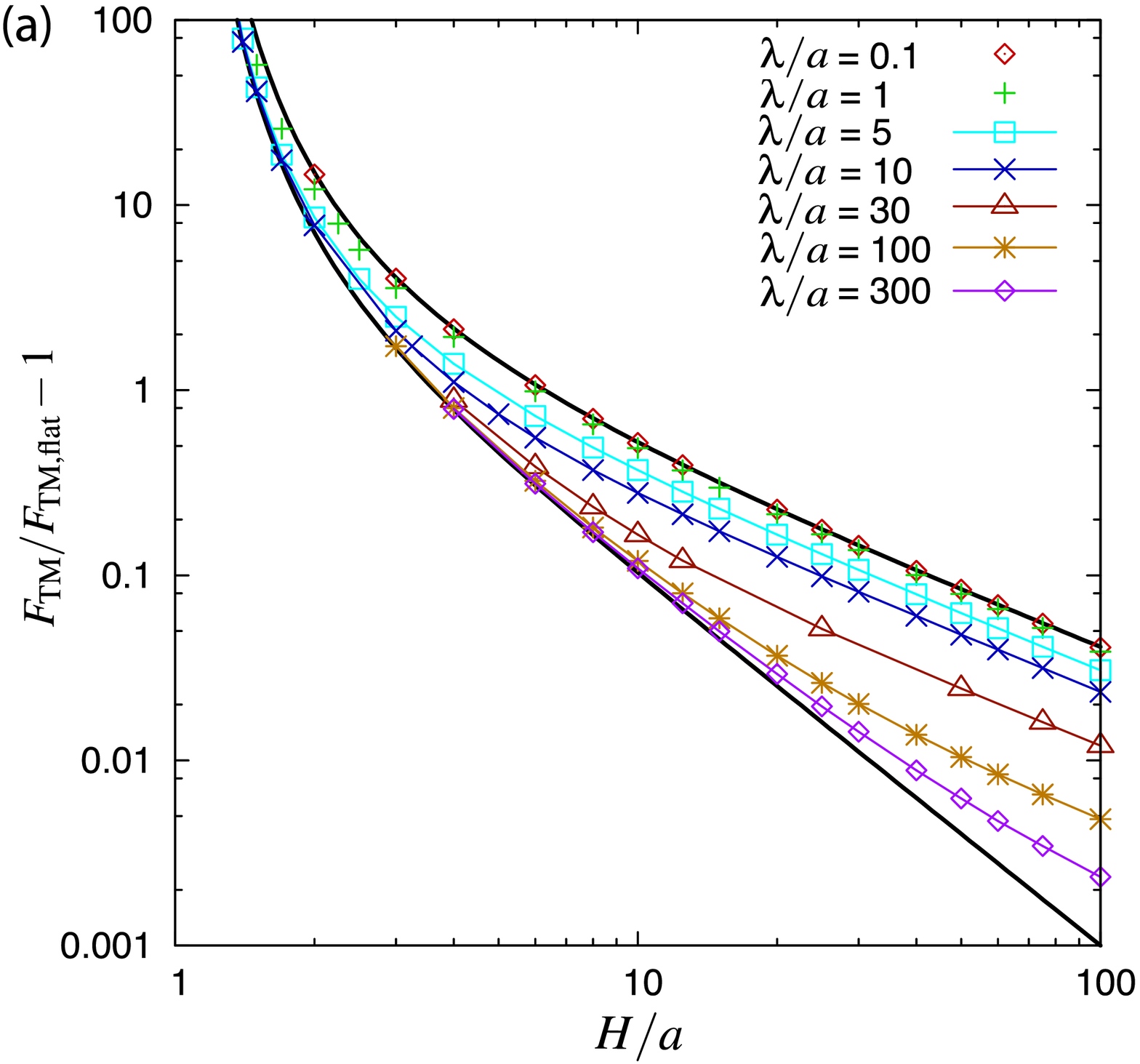}
\hfill
\includegraphics[width=0.48\linewidth]{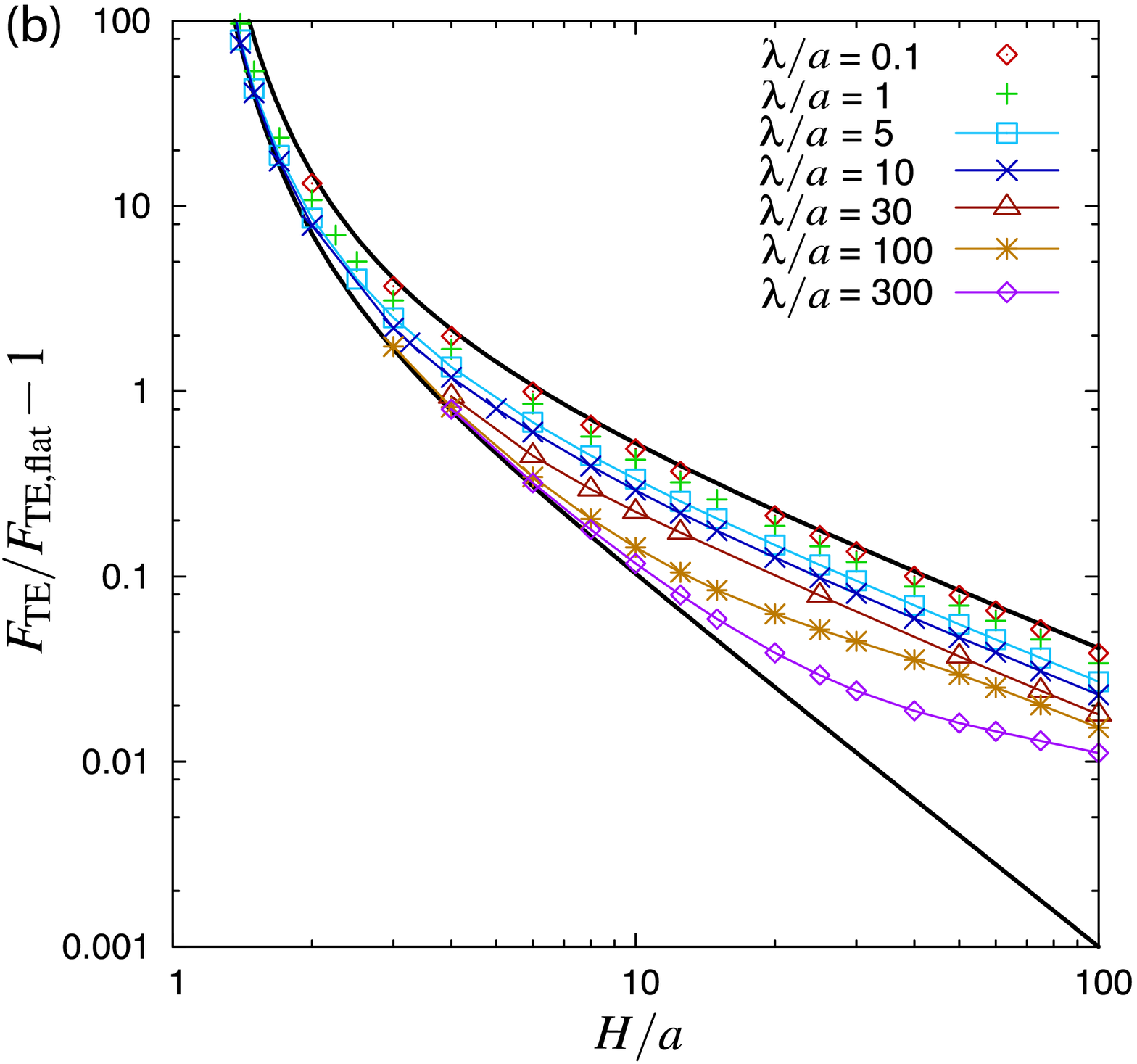}
\caption{\label{eq:fig4} (color online) Casimir force for TM modes (a) 
  and TE modes (b) as function of $H/a$ for different corrugation
  lengths $\lambda/a$. Displayed is the change of the force compared
  to the force between two flat plates,
  $F_\text{TM,flat}=F_\text{TE,flat}=-(\pi^2/480) H^{-4}$, in units of
  $F_\text{TM,flat}$ and $F_\text{TE,flat}$, respectively. The two
  bold curves enclosing the numerical data are the analytical results
  $F_0$ for $\lambda \to 0$ (upper curve) and $F_\infty$ for $\lambda
  \to \infty$ (lower curve), see text.}
\end{figure}

\begin{figure}[h!]
\includegraphics[width=0.48\linewidth]{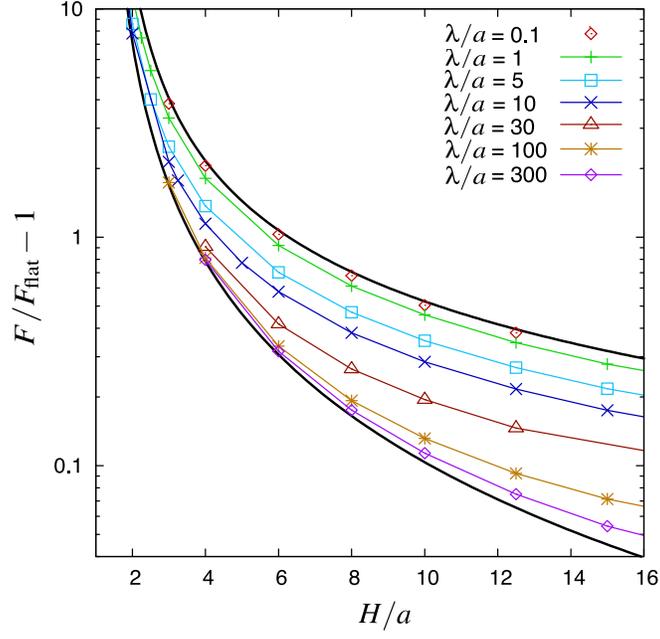}
\caption{\label{eq:fig5} (color online) Total Casimir force as sum of 
  TM and TE mode contributions in the short distance regime. Shown is
  the relative change of the force compared to the total Casimir force
  $F_\text{flat}$ between two flat plates. The data enclosing bold
  curves have the same meaning as in Fig.~\ref{eq:fig4} but for the
  total force they are now given by $2F_0$ and $2F_\infty$ due to the
  same contribution of TM and TE modes in these two limits.}
\end{figure}

\begin{figure}[h!]
\includegraphics[width=0.46\linewidth]{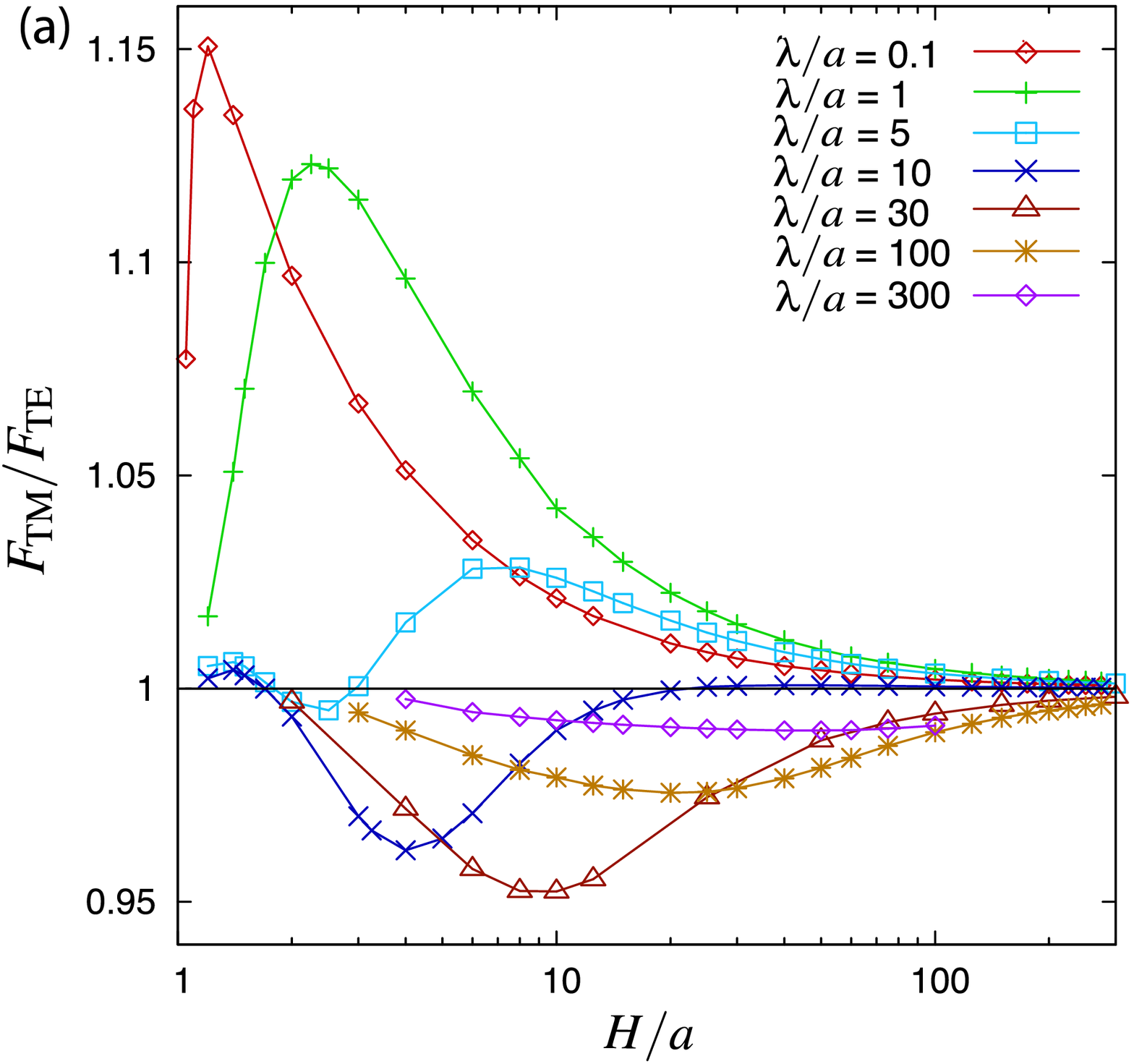}
\hfill
\includegraphics[width=0.47\linewidth]{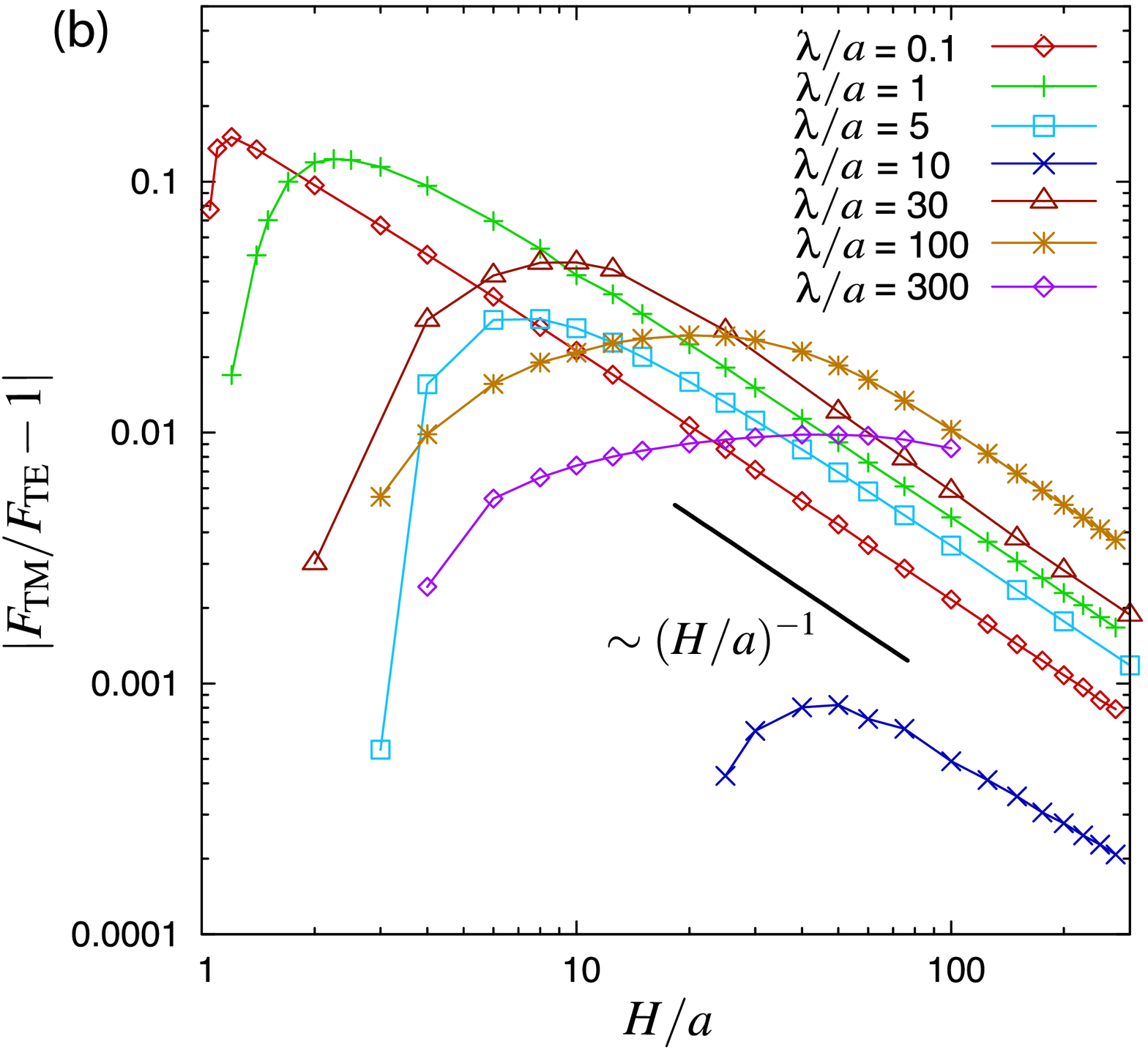}
\caption{\label{fig:ratio} (color online) (a) Ratio of Casimir force from 
  TM and TE modes as function of the plate distance $H$ for different
  corrugation lengths $\lambda$. (b) Logarithmic plot of the deviation
  of the ratio from one at large $H$.}
\end{figure}

\begin{figure}[h!]
\includegraphics[width=0.42\linewidth]{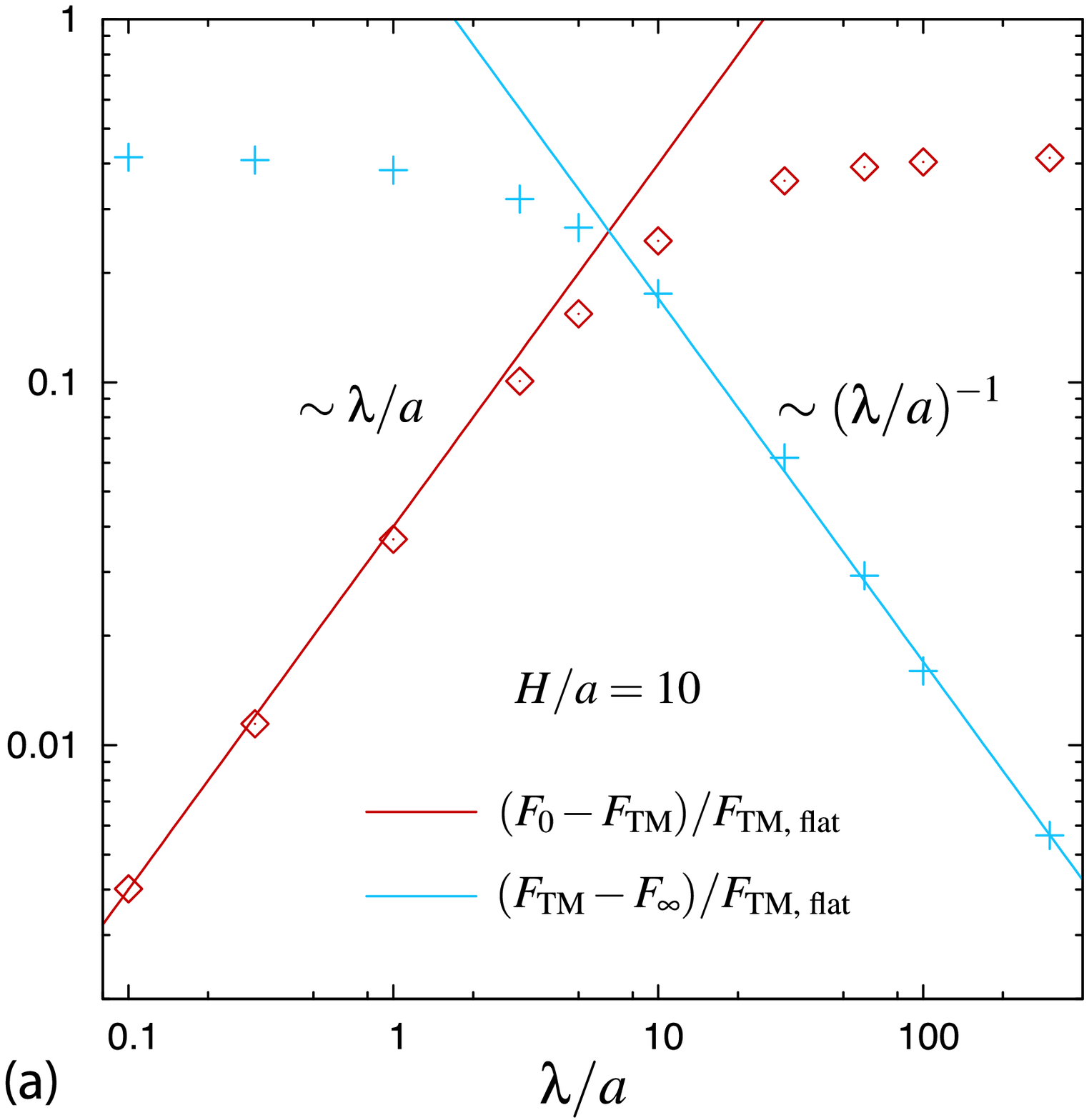}
\hfill
\includegraphics[width=0.42\linewidth]{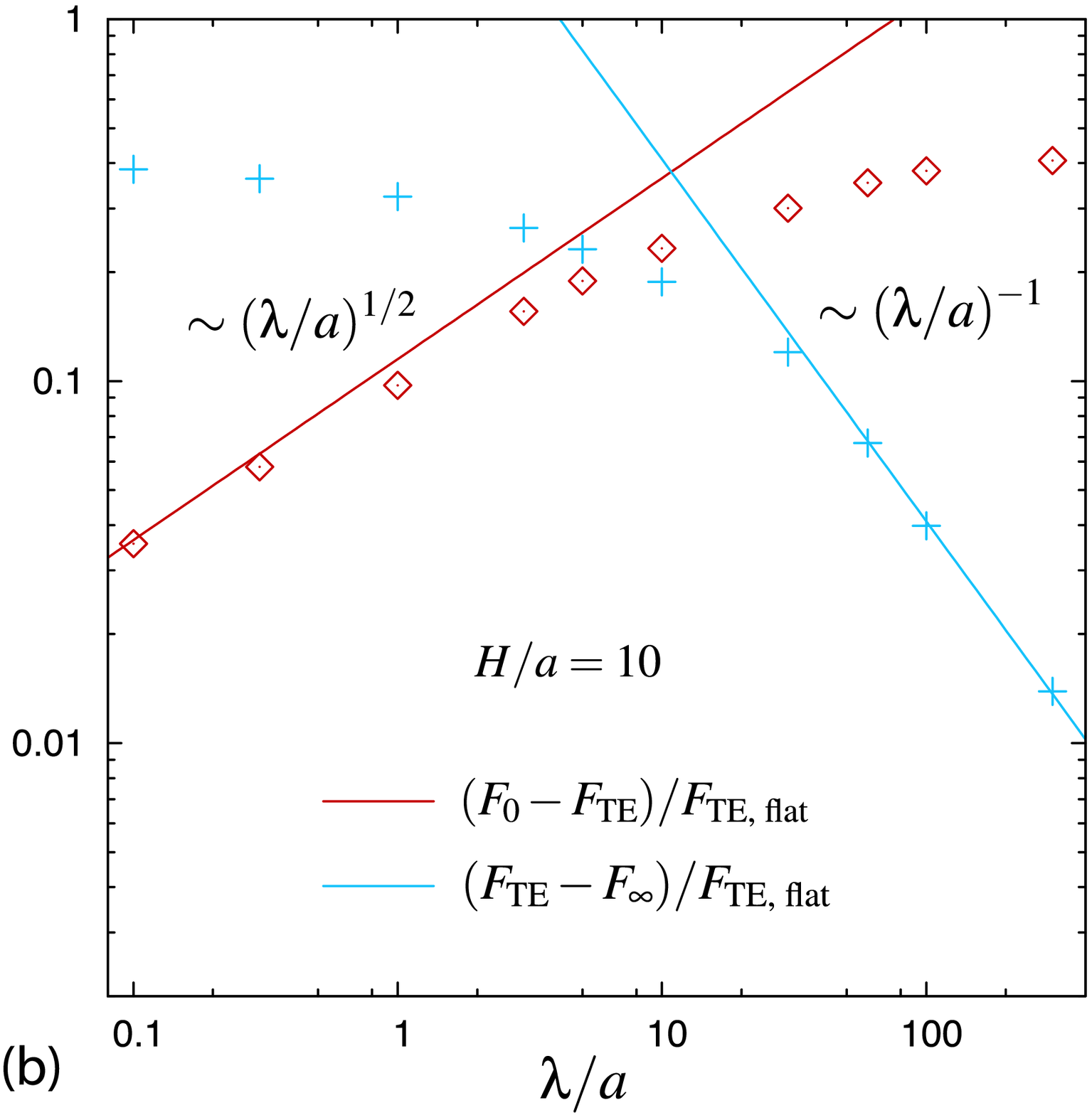}
\caption{\label{eq:fig7} (color online) Scaling of the force from 
  TM (a) and TE (b) modes close to the upper bound $F_0$ ($\lambda \to
  0$) and the lower bound $F_\infty$ ($\lambda \to \infty$) as a
  function of $\lambda/a$ at fixed mean surface distance $H=10 a$.
  Forces are measured in units of $F_\text{TM,flat}$ and
  $F_\text{TE,flat}$, respectively.}
\end{figure}

\begin{figure}[h!]
\includegraphics[width=0.42\linewidth]{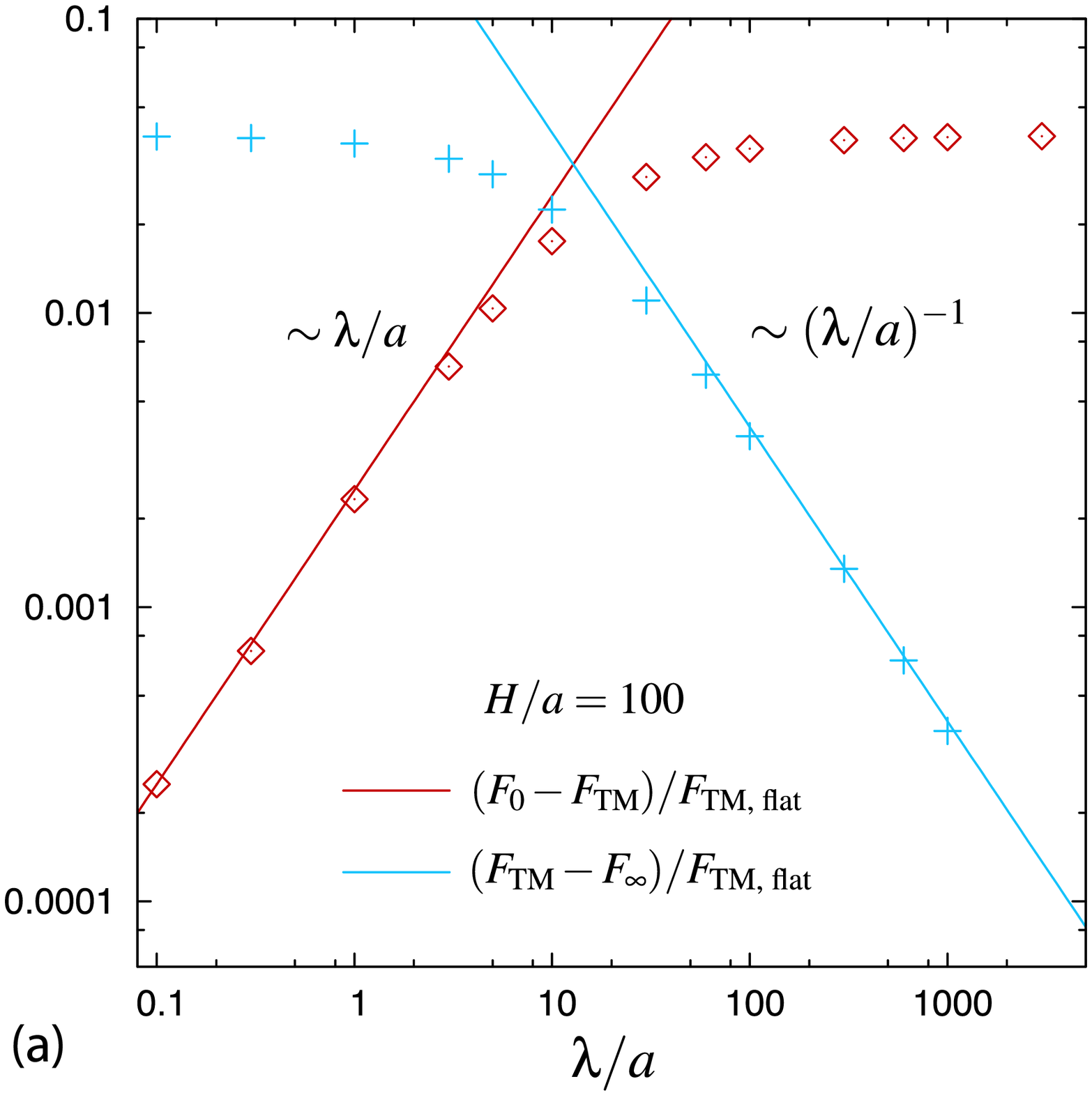}
\hfill
\includegraphics[width=0.42\linewidth]{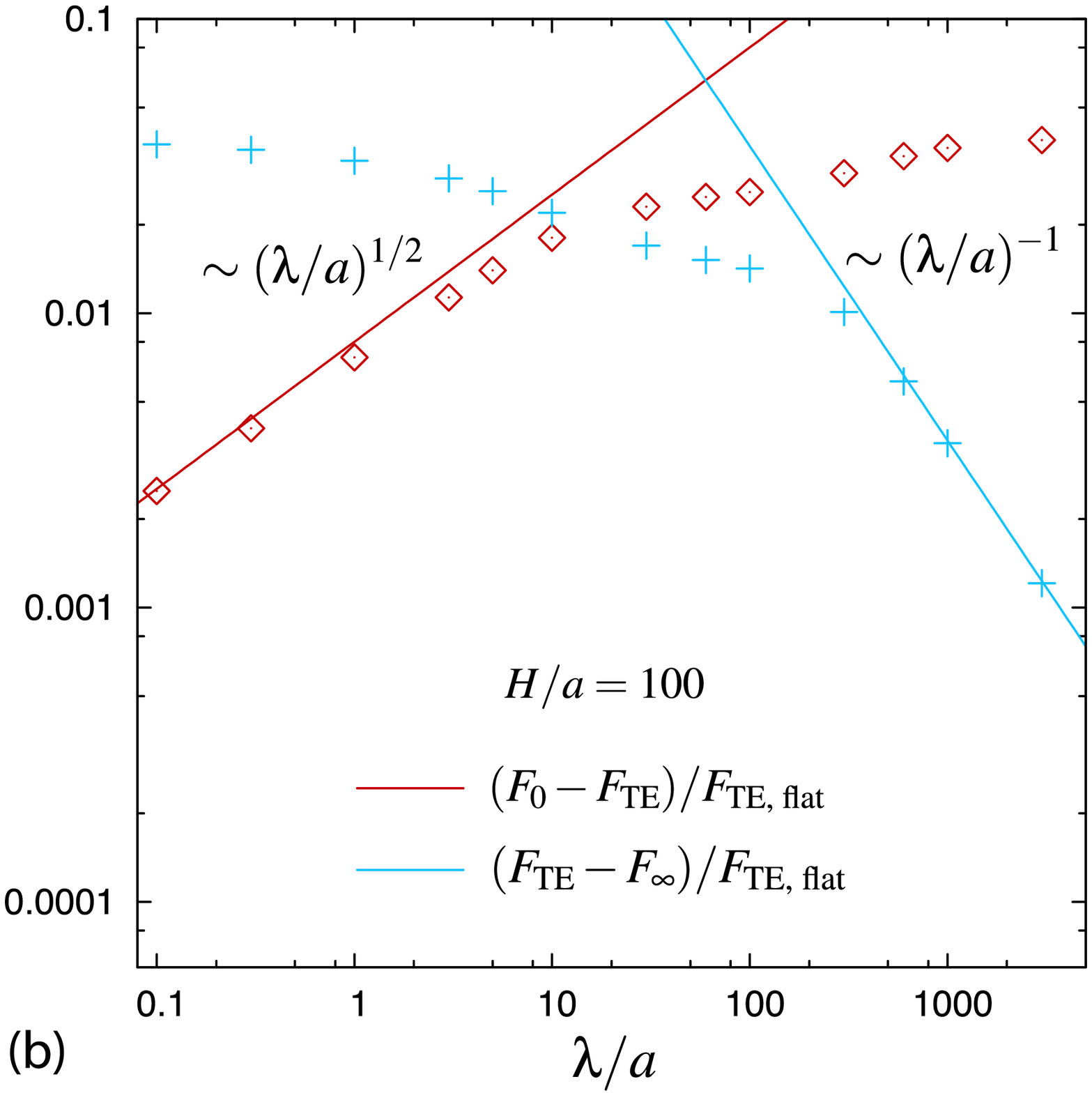}
\caption{\label{eq:fig8} (color online) Same plot as in Fig.~\ref{eq:fig7} 
  but for fixed distance $H=100 a$.}
\end{figure}

In this section we implement the non-perturbative approach of section
\ref{sec:Non-pert-pi} numerically for the rectangular corrugation of
Fig.~\ref{eq:fig1}. One has to resort to a numerical analysis here
since the function $g(q_\perp,q_1)$ cannot be obtained analytically
from the matrices of Eqs.~(\ref{eq:N_D-matrix}),
(\ref{eq:N_vN-matrix}) for arbitrary corrugation lengths $\lambda$.
The numerical procedure follows straightforwardly the computation of
the Casimir force in section \ref{sec:Non-pert-pi}. The following
implementation applies both to TM and TE modes.  At fixed order $M$,
the truncated matrix $B_{kl}$ of Eq.~(\ref{eq:matrix-B}) with $k$,
$l=-M,\ldots, M$ is calculated from the matrices $N_m$ of
Eqs.~(\ref{eq:N_D-matrix}), (\ref{eq:N_vN-matrix}). Then the matrix
$B_{kl}$ is inverted numerically to yield the function
$g_M(q_\perp,q_1)$ from Eq.~(\ref{eq:g-fct}) where the index $M$
denotes the truncation order. Notice that the derivative of $B_{kl}$
with respect to $H$ is obtained analytically and no potentially
inaccurate numerical derivatives have to be computed.  Finally, the
integration in equation (\ref{eq:force-result}) is carried out
numerically without difficulty since $g_M(q_\perp,q_1)$ decays
exponentially fast for large $q_\perp$, $q_1$.  This provides a series
of approximations $F_M$ to the Casimir force which must converge to
the exact value of the force as $M\to\infty$. From our analysis of the
flat plate geometry, see Eq.(\ref{eq:force-finite-flat}), we expect an
exponentially fast convergence $F-F_M \sim e^{-\gamma M}$ with a
coefficient $\gamma$. However, the decay coefficient $\gamma$ depends
on the geometrical lengths, and it is expected to increase with
decreasing $\lambda/H$. This type of convergence behavior we found to be
consistent with our numerical data for $F_M$. It allowed us to
extrapolate the data to obtain the Casimir force $F$. The largest $M$
for which we calculated $F_M$ was $M=10$ for small $\lambda/a=0.1$ and
$M=97$ for large $\lambda/a=300$. 

The results of our numerical analysis are as follows. If we express
the total Casimir force $F$ or the force contributions $F_\text{TM}$
and $F_\text{TE}$ from TM and TE modes, respectively, in units of the
corresponding force between two flat plates the results can be
expressed in terms of the dimensionless ratios $H/a$ and $\lambda/a$
only. The results from the extrapolation of the data for $F_M$ are
shown in Fig.\ref{eq:fig4} both for TM and TE modes and different
corrugation lengths. For both types of modes the force $F_\text{TM}$,
$F_\text{TE}$ is bounded at a fixed plate separation $H/a$ between
$F_\infty$ and $F_0$ as given by Eq.~(\ref{eq:F_infty}) and
Eq.~(\ref{eq:F_0}), respectively. For small $\lambda/a$ the upper
bound $F_0$ is approached whereas for asymptotically large $\lambda/a$
the force converges towards the lower bound $F_\infty$ which is given
by the proximity force approximation. Since the convergence towards
the lower bound $F_\infty$ becomes slower with increasing $H/a$ there
are two distinct scaling regimes for the force at a fixed corrugation
length $\lambda/a$. At small $H/a$ the relative change of the force
compared to the force between two flat plates,
$F_\text{T}/F_\text{T,flat}-1$, T=TM or TE, decays as $(H/a)^{-2}$.
After a crossover regime the relative change of the force decays at
larger $H \gg \lambda$ like $(H/a)^{-1}$, following the behavior of
the exact result $F_0$ for $\lambda \to 0$. The so far described
qualitative behavior of the force is common to both types of modes.
However, there is a clear distinction between TM and TE modes,
especially at large $\lambda/a$, as can be seen from
Fig.\ref{eq:fig4}. The force from TE modes has much more pronounced
deviation from the proximity approximation result $F_\infty$ as the TM
modes. In particular at large corrugation lengths ($\lambda/a=300$)
this can be seen clearly from our numerical data. The same behavior is
observed for the deviations from $F_0$ at small $\lambda/a$. Thus, the
force $F_\text{TE}$ appears at intermediate values of $\lambda/a$ more
strongly separated from the lower and upper bound,
cf.~Fig.~\ref{eq:fig4}(b). We will come back to this point below when
we discuss the scaling of the force with $\lambda$ close to the
bounds. Fig.~\ref{eq:fig5} shows the total Casimir force in the range
of small separations $H$.

For particular geometries like a cubic volume the Casimir force has
even a different sign for a scalar field with Dirichlet boundary
conditions (attractive force) and an electrodynamic field (repulsive
force) \cite{Boyer,Mostepanenko+88}.  Since for uniaxial plate
deformations both types of fields differ in the presence of a scalar
field with Neumann boundary conditions (TE modes) it is interesting to
study more quantitatively the difference between the two wave types.
Fig.~\ref{fig:ratio} shows the ratio $F_\text{TM}/F_\text{TE}$ of the
forces from both types of modes at different $\lambda/a$. One observes
that the ratio is peaked at a characteristic $H/a$ which depends on
$\lambda/a$. For small $H/a \to 1$ the ratio tends to one as one can
expect from the proximity force approximation which does not
differentiate between TM and TE modes. In the opposite range of large
$H/a$ again both types of modes must contribute almost equally since
the geometry approaches that of two flat plates. For the entire range
of studied corrugation lengths the ratio converges to one for large
$H/a$ according to $|F_\text{TM}/F_\text{TE}-1| \sim (H/a)^{-1}$, see
Fig.~\ref{fig:ratio}(b). However, this asymptotic behavior sets in only
beyond a crossover separation $H$ which increases with $\lambda$. At
intermediate $\lambda/a$ the ratio varies approximately between 0.95
and 1.15 in the studied range of $\lambda/a$. TM modes dominate at
$\lambda/a \lesssim 10$ and at small $H/a$ for all $\lambda/a$. The
contribution from TE waves is larger for $\lambda/a \gtrsim 10$ and
$H/a \gtrsim 2$. It is instructive to compare this behavior to
perturbative results of Ref.
\onlinecite{Emig-Hanke-Golestanian-Kardar,Emig-Hanke-Golestanian-Kardar-II}
for the geometry consisting of a {\it smooth} sinusoidally corrugated
and a flat plate. As will be explained in more detail in the next
section, the perturbative result for the later geometry yields
$F_\text{TM}/F_\text{TE}>1$ for all $\lambda/a \gg 1$ and $H/a \gg 1$,
in contrast to our results for the rectangular corrugation. This
observation suggests that the corners of the rectangular corrugation
in fact cause the slight amplification of TE modes compared to TM
waves at $\lambda/a \gtrsim 10$. One can argue that imposing for TE
modes a vanishing normal derivative on the field at the concave
corners inside the valleys of the corrugation provides a stronger
constraint on field fluctuations as compared to Dirichlet conditions
for TM modes. If the width of the valleys is decreased with $\lambda$
the two opposite corners can no longer be considered separately and
the Dirichlet condition might provide a stronger restriction. For very
small $H/a$ the main contribution to the force comes from rather short
wavelengths which should be only very weakly effected by the Neumann
conditions at the concave corners.

Finally, we consider the scaling of the force from TM and TE modes
close to lower and upper bounds $F_\infty$ and $F_0$, respectively.
Figures \ref{eq:fig7} and \ref{eq:fig8} show a logarithmic plot of
force form TM and TE modes at fixed $H=10a$ and $H=100a$, measured
relative to $F_\infty$ for large $\lambda/a$ and relative to $F_0$ for
small $\lambda/a$. At small $\lambda$ we found an interesting
qualitative difference between TM and TE modes for the scaling towards
the exact result $F_0$ for $\lambda \to 0$,
\begin{equation}
  \label{eq:F_0-scaling}
  \frac{F_0 - F_\text{TM}}{F_\text{TM,flat}} \sim \frac{\lambda}{a},
  \quad
  \frac{F_0 - F_\text{TE}}{F_\text{TE,flat}} \sim 
\left(\frac{\lambda}{a}\right)^{1/2}.
\end{equation}
For the change in the exponents we cannot present a satisfying simple
argument. In the opposite limit of large $\lambda$ the proximity
approximation result $F_\infty$ is approached linearly for both types
of modes,
\begin{equation}
  \label{eq:F_infty-scaling}
  \frac{F_\text{TM}-F_\infty}{F_\text{TM,flat}} \sim \frac{a}{\lambda},
  \quad
  \frac{F_\text{TE}-F_\infty}{F_\text{TE,flat}} \sim \frac{a}{\lambda}.
\end{equation}
As we will show in the next section, this linear decrease can be
understood in terms of geometric optics.


\section{Comparison with perturbation theory and geometric optics}

\begin{figure}[t]
\includegraphics[width=0.49\linewidth]{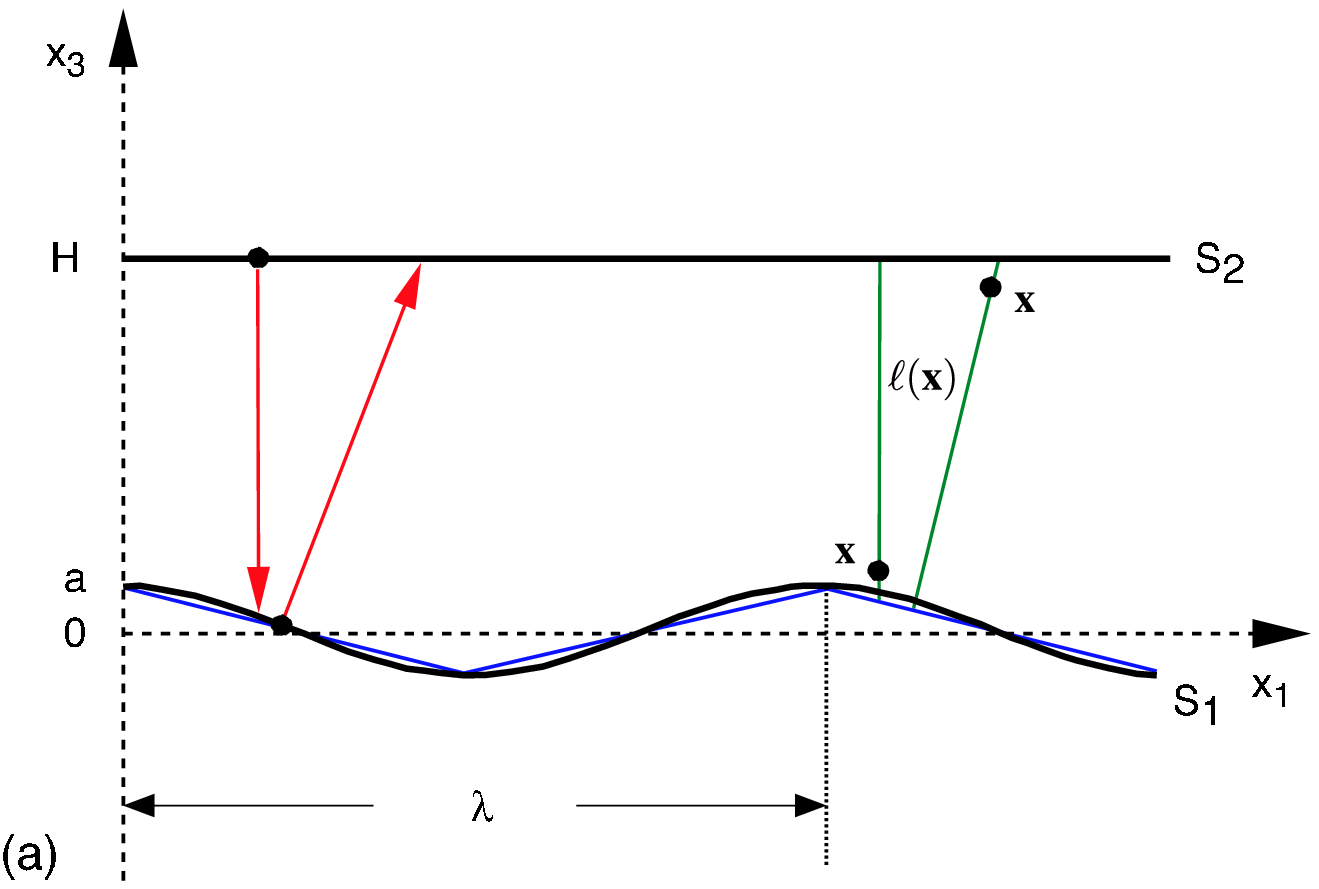}
\includegraphics[width=0.49\linewidth]{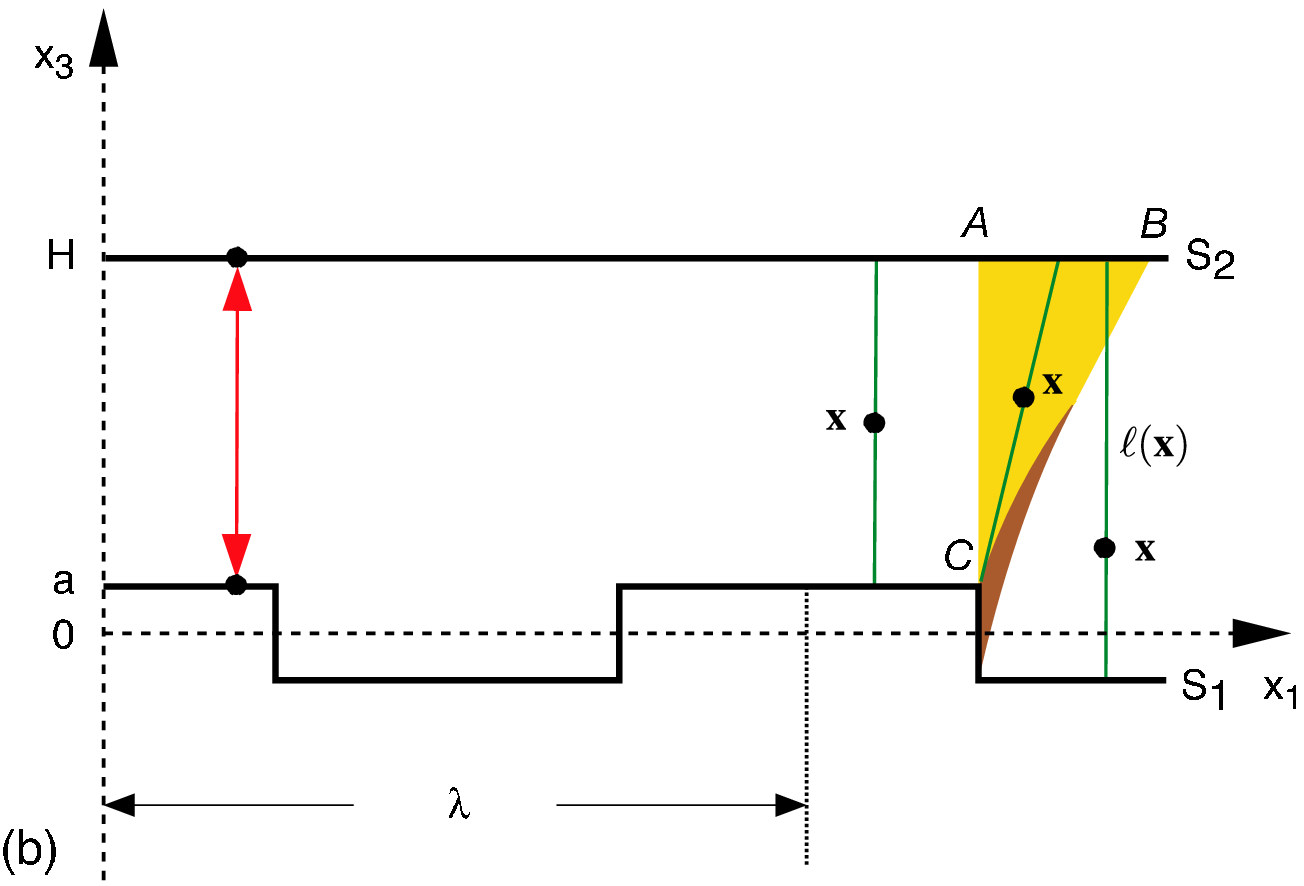}
\caption{\label{eq:fig2} (color online) Typical paths of the proximity 
  force approximation and the geometric optics approach for both
  sinusoidal (a) and rectangular corrugation (b) with $\lambda\gg a$.
  Paths with arrows denote distances which are measured normal to one
  of the surfaces as used for the proximity force approximation.
  Paths without arrows denote the shortest surface connecting paths of
  length $\ell(\xbf)$ through a point $\xbf$ located in the gap
  between the plates.}
\end{figure}

The aim of this section is to compare the numerical results of the
previous section to those which were obtained from perturbation theory
in Refs.
\onlinecite{Emig-Hanke-Golestanian-Kardar,Emig-Hanke-Golestanian-Kardar-II}
for an uniaxially and sinusoidally corrugated surface. We will show
that discrepancies in the results from the two approaches can be
qualitatively understood in terms of classical ray optics, a concept
which was introduced in Ref.  \onlinecite{jaffe} for to the
computation of Casimir interactions.  In the perturbational path
integral approach, the logarithm of the partition function is expanded
in powers of the height profile $h_1$ as
$\:\ln\Zcal=\ln\Zcal|_0+\ln\Zcal|_1+\ln\Zcal|_2+\dots\:$.  The zero
order term $\ln\Zcal|_0=\frac{\pi^2}{720} AL H^{-3}$ is the result for
flat planes.  The first order correction vanishes, $\ln\Zcal|_1=0$,
since $h_1$ is on spatial average zero, and the second order
contribution reads
\begin{equation}
\label{eq:pt-general}
\ln\Zcal|_2\:=\:\frac{\pi^2L}{240H^5}\int_{\xbf_\|}h_1^2(x_1)
-\frac{L}{4}\int_{\xbf_\|}\int_{\xbf_\|'}
K\left(\abs{\xbf_\|-\xbf_\|'}\right)
\left[h_1(x_1)-h_1(x_1')\right]^2.
\end{equation}
where $K(\abs{\xbf_\|-\xbf_\|'})$ denotes a response kernel which has
contributions from both TM and TE modes and was obtained in Ref.
\onlinecite{Emig-Hanke-Golestanian-Kardar-II}. The second term is only
finite for a smooth profile $h_1(x_1)$ since the kernel has a
singularity $\sim \abs{\xbf_\|-\xbf_\|'}^{-3}$. Thus for a rectangular
corrugation with $\int_{\xbf_\|} [h_1(x_1)-h_1(x_1+x_1')]^2 \sim
|x_1'|$ for $|x_1'| < \lambda/4$ the perturbative result diverges due
to the presence of sharp edges in the surface profile. In contrast,
for a sinusoidal profile with $h_1(x_1)=a\cos(2\pi x_1/\lambda)$ one
has $\int_{\xbf_\|} [h_1(x_1)-h_1(x_1+x_1')]^2 \sim x^{'2}_1$ and the
divergence of the kernel is compensated.  For this reason, we compare
our numerical results for the rectangular corrugation to the
perturbative results for a sinusoidal profile
\cite{Emig-Hanke-Golestanian-Kardar,Emig-Hanke-Golestanian-Kardar-II}.
This will allow us to study the influence of edges on the Casimir
interaction.  Perturbation theory yields for the total Casimir force
of the sinusoidal geometry of Fig.\ref{eq:fig2}(a) the result
\begin{equation}
\label{eq:PT_force}
F\:=\:F_\text{flat}\left[
1+
\tilde G\left(\frac{H}{\lambda}\right)
\left(\frac{a}{H}\right)^2+\Ocal(a^3)\right],
\end{equation}
with the function parameter free $\tilde
G(u)=\frac{480}{\pi^2}[5G(u)-uG'(u)]$ where
$G(u)=G_\text{TM}(u)+G_\text{TE}(u)$ has contributions from TM and TE
modes; for the explicit form of $G(u)$ see
Ref.~\onlinecite{Emig-Hanke-Golestanian-Kardar-II}. For comparison
with our numerical results the limits of small and large $H/\lambda$
are of particular interest. From an expansion of $\tilde G(u)$ one
obtains the asymptotic expressions
\begin{equation}
\label{eq:G_total}
\frac{F}{F_\text{flat}}-1\:=\:\left\{
\begin{array}{ll}
\frac{8\pi}{3}\frac{a}{\lambda}\frac{a}{H} & \mbox{for $\lambda\ll H$}
\\[5pt]
5\left(\frac{a}{H}\right)^2+\left(\frac{4\pi^2}{3}-20\right)
\left(\frac{a}{\lambda}\right)^2
 & \mbox{for $\lambda\gg H$}
\end{array}\right. .
\end{equation}
In both limits the results are valid only if $a \ll \lambda$. In the
limit of small $\lambda/a$ there is a divergence $\sim a/\lambda$ in
the perturbative result which reflects the above mentioned divergence
in Eq.~(\ref{eq:pt-general}) for rectangular corrugations with
vertical segments.  This singularity does not appear in our numerical
results of the previous section; it is a characteristic feature of
perturbation theory. In the following comparison we consider only the
case $\lambda\gg a$. Eq.~(\ref{eq:G_total}) suggests for large plate
separations $H \gg \lambda$ a decay of the excess force from the
corrugation $\sim a/H$ and for small $H/\lambda$ a decay $\sim
(a/H)^2$. The scaling behavior is in agreement with our observations
for the rectangular corrugation as demonstrated by Fig.\ref{eq:fig4}.
However, the latter Figure also shows that for smaller $\lambda/a
\lesssim 10$ the scaling regime with a decay $\sim (a/H)^2$ does not
exist.

Next, we will compare the perturbative results of
Eq.~(\ref{eq:G_total}) with our numerical results for the deviation of
the actual Casimir force from the proximity force approximation (PA),
$(F-F_\text{PA})/F_\text{flat}$, where $F_\text{PA}$ is the force
obtained from the PA. This approximation does no distinguish between
the two types of modes and thus for the rectangular corrugation one
has $F_\text{PA}=2 F_\infty$ with $F_\infty$ given by
Eq.~(\ref{eq:F_infty}).  In general, for deformed surfaces the PA is
ambiguous \cite{jaffe} since the pairs of small parallel surface
elements can be chosen to be parallel to either surface so that the
local plate distance is measured either normal to $S_1$ or normal to
$S_2$ as indicated by the arrows of Fig.\ref{eq:fig2}. We emphasize
that this ambiguity does not arise for the rectangular corrugation.
For smooth surfaces with finite curvature like a sinusoidal
corrugation the PA result depends on the reference plate.  If one
measures the local distance perpendicular to the flat surface, as it
is most common, one obtains for the Casimir energy per surface area
\begin{equation}
  \label{eq:PA-flat-based}
  \Ecal_\text{PA}
=\frac{1}{A}\int_{S_2}dS\,\Ecal_\text{flat}[H-h_1(x_1)],
\end{equation}
but if the local distances are chosen perpendicular to the corrugated
plate, one has
\begin{equation}
  \label{eq:PA-corr-based}
\Ecal_\text{PA,corr}=\frac{1}{A}\int_{S_1}dS\,
\Ecal_\text{flat}\left[(H-h_1(x_1))\sqrt{1+(h_1'(x_1))^2}\right],
\end{equation}
where $\Ecal_\text{flat}(H)=-\frac{\pi^2}{720}H^{-3}$ is the Casimir
energy per surface area for two flat surfaces. For a sinusoidal
corrugation the integrals over the surfaces can be computed
perturbatively in $a$. This yields for large $\lambda$ the difference
between the force $F$ from perturbation theory [Eq.(\ref{eq:G_total})]
and the PA force, based on the flat and the corrugated plate,
respectively,
\begin{equation}
\label{eq:F_lim}
\frac{F-F_\text{PA}}{F_\text{flat}}\:=\:
\left(\frac{4\pi^2}{3}-20\right)\left(\frac{a}{\lambda}\right)^2\quad,\qquad
\frac{F-F_\text{PA,corr}}{F_\text{flat}}\:=\:
\left(\frac{10\pi^2}{3}-20\right)\left(\frac{a}{\lambda}\right)^2.
\end{equation}
The essential result is that the perturbatively obtained force
approaches the PA approximation like $(a/\lambda)^2$ for large
$\lambda$ which has to be compared to the $a/\lambda$ decay seen in
our numerical results for the rectangular corrugation,
cf.~Figs.~\ref{eq:fig7} and \ref{eq:fig8}. Thus the deviation from the
PA is stronger for the rectangular corrugation than for the sinusoidal
profile, presumably due to sharp edges. Before we give a simple
physical argument for the variation of the decay exponent let us
compare the amplitudes in Eq.~(\ref{eq:F_lim}). If we chose the PA to
be based on the flat plate, the amplitude is negative, and the force
$F_\text{PA}$ is {\it not} a lower bound to the force at a fixed
$H/a$, in disagreement with our observation for a rectangular
corrugation.  The corrugated surface based PA in contrast yields a
positive amplitude. We expect that also for a sinusoidal corrugation
the actual force is monotonous in $\lambda/a$ at fixed $H/a$, assuming
its minimal value for $\lambda/a \to \infty$. The change of sign is
just an other manifestation of the ambiguity in the proximity
approximation. The observation that the actual Casimir force is
located between the flat and the curved surface based PA was also made
for a plane plate--sphere geometry recently \cite{Gies+03}.

In order to understand the dependence of the exponent for the scaling
towards the PA limit on the shape of the corrugation it is instructive
to consider classical ray optics. Such an approach was recently
applied to the calculation of Casimir interactions \cite{jaffe}. Since
this approach does not take into account diffraction it is limited to
deformations where the radii of curvature are large compared to the
smallest distance between the surfaces. But still, geometric optics
allow for a better description of Casimir forces than the conventional
proximity force approximation. By considering instead of all actual
optical paths only the {\it shortest} paths, Jaffe and Scardicchio
proposed an ``optimal'' proximity approximation for scalar field
fluctuations subject to Dirichlet boundary conditions \cite{jaffe}.
It can be also applied to electromagnetic fields. Consider a position
$\xbf$ in the vacuum space between the plates, and denote by $\ell(\xbf)$
the length of the shortest optical ray between the plates through that
point. Fig.\ref{eq:fig2} shows typical paths for the two types of
corrugations we consider here. The Casimir energy in this optical
approximation can then be written as
\begin{equation}
\label{eq:optical_integral}
\frac{\Ecal_\text{opt}}{\Ecal_\text{flat}}\:=\:
\int d^2 \xbf_\|\int_{h_1(x_1)}^H dx_3\,
\frac{H^3}{A\ell^4(\xbf_\|,x_3)},
\end{equation}
where the integral runs over the total space between the surfaces.

First, we apply this approach to the sinusoidal profile, see
Fig.\ref{eq:fig2}(a).  For simplicity, we replace the sinusoidal
profile by a piecewise linear profile, cf.~Fig.\ref{eq:fig2}(a), which
is a good approximation in the limit $a\ll\lambda$ considered here.
Then we have to determine $\ell(\xbf)$ for each position between the
plates for this simpler profile. Since the exact value for
$\ell(\xbf)$ is difficult to evaluate, we consider the two cases where
the position is close to one of the two surfaces and then assume a
linear interpolation between the two length for $\ell(\xbf)$ at
arbitrary $\xbf$ in the gap between the plates. If $\xbf$ is very
close to the deformed surface $S_1$ the shortest paths is
perpendicular to the flat surface $S_2$. Contrary, if $\xbf$ is
located close to the flat surface $S_2$ the shortest ray is
perpendicular to the deformed surface $S_1$.  With the so obtained
approximative lengths $\ell(\xbf)$ we obtain from
Eq.~(\ref{eq:optical_integral}) by expansion in $a/H$ for the
correction to the flat surface based proximity approximation the
scaling behavior
\begin{equation}\label{eq:F_lim-2}
\frac{F_\text{opt}-F_{\infty}}
{F_\text{flat}} \: \sim \:\left(\frac{a}{\lambda}\right)^2.
\end{equation}
Thus the optical approach nicely reproduces the correct scaling of the
corrections to the proximity approximation at large $\lambda$, in
agreement with the perturbative result of Eq.~(\ref{eq:F_lim}). 

In order to examine the role of edges for deviations from the
proximity approximation, we apply the optical approach also to the
rectangular corrugation of Fig.\ref{eq:fig2}(b). For this geometry the
shortest paths are easily identified. Except for positions located in
an almost triangular shaped region (composed of the two shaded regions
of Fig.\ref{eq:fig2}(b) the paths are just perpendicular to both
surfaces. Thus the deviation from the proximity approximation is
caused by paths through points which are located inside the shaded
region. These paths run either to corner $C$ of the surface (larger
region) or to the vertical surface segment (smaller region).  For
sufficiently large $\lambda$ the regions from adjacent edges do not
overlap and can be treated independently. Furthermore, since the ratio
of the area of the larger shaded region formed by the triangle $ABC$
and the area of the smaller shaded region bounded by the vertical
surface segment scales like $\sim (H/a)^2$ one has to consider only
the triangle $ABC$ for the evaluation of
Eq.~(\ref{eq:optical_integral}) in the limit $a/H \ll 1$.  This gives
\begin{equation}\label{eq:F_lim-3}
\frac{F_\text{opt}-F_{\infty}}
{F_\text{flat}}\:\sim\:\sqrt{\frac{a}{H}}\frac{a}{\lambda}.
\end{equation}
This result is in agreement with the scaling behavior we have observed
in our non-perturbative approach for the rectangular profile, see
Figs.\ref{eq:fig7} and \ref{eq:fig8}, and
Eq.~(\ref{eq:F_infty-scaling}). We conclude that the analysis of the
{\it shortest} optical paths explains the observed dependence of the
Casimir force on the surface shape close the proximity force limit
$\lambda \gg H$.

Finally, we consider the ratio $F_\text{TM}/F_\text{TE}$ of the force
contributions from TM and TE modes. In perturbation theory one obtains
from the separate contributions of the two types of modes to the
result of Eq.~(\ref{eq:PT_force}) the low $a$ expansion
\begin{equation}
\label{eq:ratio_TM_TE}
\frac{F_\text{TM}}{F_\text{TE}}\:=\:
1+\frac{8\pi}{3}\frac{a}{\lambda}\frac{a}{H},
\end{equation}
which is valid if both $H \gg \lambda$ and $\lambda \gg a$. Thus for
sinusoidal corrugations the force has always larger contributions from
TM modes at asymptotically large $H$, in contrast to our numerical
results for rectangular corrugations, cf.~Fig.\ref{fig:ratio}(a). We
argued in the previous section that edges might cause the
amplification of TE mode contributions. However, the convergence of
the ratio to one for large $H$ turns out be insensitive to the shape
of the corrugations.  Our numerical results agree perfectly over the
full range of studied $\lambda/a$ with perturbation theory in that the
ratio decays like $a/H$ to one, see Fig.\ref{fig:ratio}(b). For small
$\lambda/a \to 0$ the amplitude in no longer given by
Eq.~(\ref{eq:ratio_TM_TE}) but saturates at a finite value which
decreases with $\lambda$ since for $\lambda \to 0$ the reduced
distance argument of section \ref{sec:small-lam} implies equal
contributions from both types of modes.


\section{Conclusions and further applications}

In this paper we have developed a non-perturbative method to compute
Casimir interactions in periodic geometries. This approach is based on
a path integral quantization of the electromagnetic field subject to
ideal metal boundary conditions. The so obtained effective action for
the Casimir interaction is transformed to a representation which is
adapted to periodic geometries and allows for an efficient numerical
computation of the force between macroscopic objects.  In particular,
the approach allows us to compute the Casimir force between surfaces
with {\it strong} periodic deformations and edges. For uniaxial
deformations the electromagnetic field can be decomposed into two
scalar fields which are subject to Dirichlet and Neumann boundary
conditions, respectively. This enables us to study qualitative
differences in the geometry dependence of the Casimir interaction for
scalar fields with different boundary conditions. Applications of the
latter case range from thermal fluctuations in superfluids to liquid
crystals \cite{likar-letter,likar} which can be described by a scalar
field.  Path integral quantization in the presence of boundaries has
been previously applied to perturbative calculations of Casimir
interactions between static and dynamic deformed manifolds in the
context of both thermal \cite{likar-letter,likar} and quantum
fluctuations
\cite{golkar-letter,golkar,Emig-Hanke-Golestanian-Kardar,Emig-Hanke-Golestanian-Kardar-II}
of the confined field. However, all these computations were restricted
to slightly deformed surfaces and edges were excluded. While a number
of qualitative predictions of perturbation theory are confirmed by our
approach even for strong deformations, we find novel non-perturbative
effects which were unaccessible previously.

As an explicit example, we calculated by the Casimir interaction
between a flat and a rectangular corrugated plate with edges,
including the case of large deformation amplitudes.  Arbitrary
periodic profiles can be treated by our approach as well by Fourier
transforming the kernel of the effective action numerically and then
applying the same technique we used here for the rectangular
corrugation.  We could confirm the perturbatively predicted existence
of two different scaling regimes for the deformation induced part of
the interaction as a function of the mean plate separation $H$.
However, we also find that for small corrugation lengths only the
large $H$ scaling regime exists. We demonstrate by explicit
calculations that in the limit of very small corrugation lengths the
force can be obtained as the interaction of two flat surfaces with a
reduced distance. At very large corrugation length and small $H$ we
find that the force approaches the result of the proximity force
approximation. Our approach also allowed for a precise computation of
the scaling of the force close to the limits of small and large
corrugation length which provide an upper and lower bound,
respectively, to the force. In both cases we find power law scaling
with $\lambda/a$, rendering corrections to proximity approximation in
general large. The exponents of these power laws depend on the type of
modes (transversal electric or magnetic) for small corrugation length.
At large corrugation length we find an interesting dependence of the
exponents on generic features of the corrugations. By comparison with
perturbation theory for a sinusoidal corrugation we find that edges
induce a slower decay towards the prediction of the proximity
approximation as compared to smooth profiles. We could explain this generic
behavior in terms of classical optical paths.

Our non-perturbative method can be applied to a number of other
interesting situations. Since the path integral technique can be used
in arbitrary dimensions of the embedding space and the surfaces, our
method can be also used in this general cases.  In this paper we
focused on uniaxial deformations. Two directional corrugations can
also be treated by our method by applying it to the full
electromagnetic gauge field without splitting into TM and TE modes.
The latter case could help to understand the possibility of repulsive
forces since plates with two directional corrugations form at short
distances cavities, i.e., geometrical shapes similar to a sphere for
which a repulsive ``force'' is expected \cite{Boyer}. At short plate
separations, material properties become in general important for the
interaction. These effects can be also described by a path integral
approach with non-local boundary conditions \cite{Buescher+03},
enabling the application of the methods developed here. For two
corrugated surfaces, the existence of a lateral Casimir force has been
predicted and computed by perturbative techniques
\cite{golkar-letter,Emig-Hanke-Golestanian-Kardar-II}. It would be
interesting to study the effect of {\it strong} corrugations and edges
on the lateral Casimir effect by our method. For the dynamic Casimir
effect the surfaces are {\it dynamically} deformed which leads for
oscillations in time again to corrugated surfaces in Euclidean space
but now along imaginary time. Our results thus imply different
behavior at small and large frequencies.


\appendix



\section{Fourier transform of the rectangular corrugation model}

\begin{figure}[t]
\includegraphics[width=0.5\linewidth]{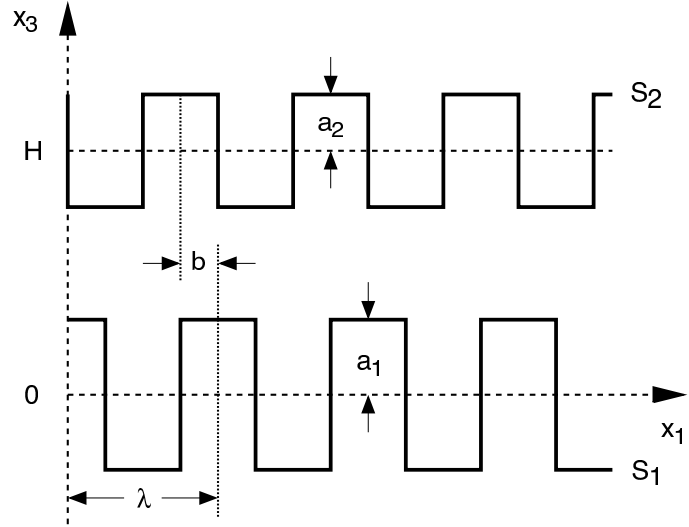}
\caption{\label{fig1} Two rectangular corrugated plates with the same
  wavelength $\lambda$ but different amplitudes $a_1$ and $a_2$ and a
  lateral shift of $b$.  The plates are translationally invariant
  along the $x_2$ direction.}
\end{figure}

We calculate the Fourier transformed matrices $\til{\Mcal}$
for Dirichlet and Neumann boundary conditions for a slightly
more general geometry with two corrugated plates.
Both plates are assumed to have a rectangular corrugation
profile with the same wavelength $\lambda$,
but with different amplitudes $a_1$ and $a_2$.
This geometry is depicted in Fig.\ref{fig1}, and the geometry of
the system discussed in section III is
obtained by simply setting the amplitude of the second plate to zero.
The reason to perform this calculation here is that it is
more transparent than the calculation which assumes one
corrugated and one flat plate. In addition, we allow the plates
to have a lateral displacement $b$.

We start with the matrix for Dirichlet boundary conditions, cf.
Eq.~(\ref{eq:Dirichlet-matrix-M}).  Performing first the Fourier
transformation with respect to $\xbf_\perp=(x_0,x_2)$, we have
\begin{equation}\label{eq:TM-Matrix}\begin{split}
\til{\Mcal}_\text{D}^{\alpha\beta}\left(\pbf,\qbf\right) &=
\int_{\xbf_\perp}\!
\int_{\ybf_\perp}\!\int_{x_1}\!\int_{y_1}
e^{i\pbf_\perp\!\cdot\xbf_\perp+i\qbf_\perp\!\cdot\ybf_\perp}
e^{ip_1x_1+iq_1y_1}\,\Gcal\left(\xbf_\perp\!-\ybf_\perp,x_1\!-y_1;
h_\alpha(x_1)-h_\beta(y_1)+H(\delta_{\alpha2}-\delta_{\beta2})\right)\\[5pt]
&=(2\pi)^2\delta^{(2)}(\pbf_\perp\!+\qbf_\perp)\int_{x_1}\int_{y_1}\int_{p_1'}
e^{i(p_1-p_1')x_1+i(q_1+p_1')y_1}\,
\frac{e^{-\sqrt{p_\perp^2\!+p_1'^2}\,
\abs{h_\alpha(x_1)-h_\beta(y_1)+H(\delta_{\alpha2}-\delta_{\beta2})}}}
{2\sqrt{p_\perp^2\!+p_1'^2}}.
\end{split}\end{equation}
To evaluate this last expression analytically,
it is necessary to find a simplified expression for the
dependence of the second exponential term on $x_1$ and $y_1$.
At this point, the use of piecewise constant profiles
for the material plates becomes crucial:
Since $h_\alpha=\pm a_\alpha$,
for $\alpha=\beta$ we can write
\begin{equation}\label{eq:trick1}
e^{-\til{p}\abs{h_\alpha(x_1)-h_\alpha(y_1)}}\:=\:e^{-a_\alpha\til{p}}
\big[\ch(a_\alpha\til{p})+
a_\alpha^{-2}\,h_\alpha(x_1)h_\alpha(y_1)\sh(a_\alpha\til{p})\big]\,.
\end{equation}
Similarly, for $\alpha\not=\beta$, we get
\begin{equation}\label{eq:trick2}
e^{-\til{p}
\abs{h_\alpha(x_1)-h_\beta(y_1)+H(\delta_{\alpha2}-\delta_{\beta2})}}\:=\:
e^{-\til{p}H}\,
\big[\ch(a_\alpha\til{p})-(-1)^\alpha a_\alpha^{-1}h_\alpha(x_1)
\sh(a_\alpha\til{p})\big]
\big[\ch(a_\beta\til{p})-(-1)^\beta a_\beta^{-1}h_\beta(x_1)
\sh(a_\beta\til{p})\big].
\end{equation}
To keep the notation short, we introduced
$\til{p}=\sqrt{p_\perp^2\!+p_1'^2}$.  Now, we insert the Fourier
series expression for $h_\alpha$ given by
\begin{equation}\label{eq:series}
h_\alpha(x_1)\:=\:\frac{2a_\alpha}{\pi}
\sum_{n=-\infty}^\infty \frac{(-1)^{n-1}}{2n-1}\,
e^{\frac{2\pi i}{\lambda}(2n-1)(x_1+\delta_{\alpha2}b)}
\end{equation}
into the rhs of Eqs.~(\ref{eq:trick1}) and (\ref{eq:trick2}).
Then, inserting those into Eq.~(\ref{eq:TM-Matrix}),
the remaining integrals over $x_1,y_1$ and $p_1'$ can easily be performed.
This yields the periodic formula
\begin{equation}\label{eq:structure}
\til{\Mcal}_\text{D}\left(\pbf,\qbf\right)
\:=\:(2\pi)^3\delta^{(2)}\left(\pbf_\perp\!+\qbf_\perp\right)\,
\sum_{m=-\infty}^\infty
\delta\left(p_1+q_1+2\pi m/\lambda\right)\,
N_{\text{D},m}\left(q_\perp,q_1\right)
\end{equation}
with the matrices
\begin{equation}
N_{\text{D},m}\left(q_\perp,q_1\right)\:=\:
\left(
\begin{array}{cc}
A^{\text{D}}_{m,1}\left(q_\perp,q_1\right) &
B^{\text{D}}_{m,12}\left(q_\perp,q_1\right)\\[5pt]
\gamma^mB^{\text{D}}_{m,21}\left(q_\perp,q_1\right) &
\gamma^mA^{\text{D}}_{m,2}\left(q_\perp,q_1\right)
\end{array}\right)\:+\:\delta_{m0}\left(\begin{array}{cc}
\frac{1}{4q}(1+e^{-2a_1q}) & \frac{e^{-qH}}{2q}\ch(a_1q)\ch(a_2q)\\[5pt]
\frac{e^{-qH}}{2q}\ch(a_1q)\ch(a_2q) & \frac{1}{4q}(1+e^{-2a_2q})
\end{array}\right)
\end{equation}
for $m$ even, and
\begin{equation}
N_{\text{D},m}\left(q_\perp,q_1\right)\:=\:
\left(
\begin{array}{cc}
0 & C^{\text{D}}_{m,12}\left(q_\perp,q_1\right)\\[5pt]
C^{\text{D}}_{m,21}\left(q_\perp,q_1\right) & 0
\end{array}\right)
\end{equation}
for $m$ odd. The entries of the matrices are given as follows
\begin{eqnarray}
A^{\text{D}}_{m,\alpha}\left(q_\perp,q_1\right) &=&
\frac{(-1)^{\frac{m}{2}}}{\pi^2}\sum_{k=-\infty}^\infty\frac{1}{(m-2k+1)(2k-1)}
\frac{e^{-2a_\alpha\til{q}_{2k-1}}-1}{\til{q}_{2k-1}}\,,\\[5pt]
B^{\text{D}}_{m,\alpha\beta}\left(q_\perp,q_1\right) &=&
2\,\frac{(-1)^\frac{m}{2}}{\pi^2}\sum_{k=-\infty}^\infty\frac{
\gamma^{(2k-1)(\delta_{\beta2}-\delta_{\alpha2})}}
{(m-2k+1)(2k-1)}
\frac{e^{-\til{q}_{2k-1}H}}{\til{q}_{2k-1}}
\sh(a_\alpha\til{q}_{2k-1})\sh(a_\beta\til{q}_{2k-1})\,,
\end{eqnarray}
and
\begin{equation}
C^{\text{D}}_{m,\alpha\beta}\left(q_\perp,q_1\right)\:=\:
\frac{(-1)^\frac{m+1}{2}}{m\pi}\left[
(-1)^\alpha\gamma^{m\delta_{\alpha2}}\frac{e^{-qH}}{q}
\sh(a_\alpha q)\ch(a_\beta q)+
(-1)^\beta\gamma^{m\delta_{\beta2}}
\frac{e^{-\til{q}_mH}}{\til{q}_m}
\sh(a_\beta\til{q}_m)\ch(a_\alpha\til{q}_m)\right]\,,
\end{equation}
where the phase factor $\gamma=e^{\frac{2\pi i}{\lambda}b}$ was
introduced.  We note that the off diagonal entries
$B^\text{D}_{m,\alpha\beta}$ and $C^\text{D}_{m,\alpha\beta}$
implicitly depend on $b$ through $\gamma$.  Furthermore,
$\til{q}_n=\sqrt{q_\perp^2+(q_1+2\pi n/\lambda)^2}$ was introduced,
which implies $q\equiv\til{q}_0$.  If $a_2=0$, the matrices
$N_{\text{D},m}$ have the symmetry
$N_{\text{D},m}\left(q_\perp,-q_1\right)=N_{\text{D},-m}\left(q_\perp,q_1\right)$,
and analogously for the Neumann matrices $N_{\text{N},m}$ which we
used in Sec.~II. We remark that this symmetry is no longer valid for
either type of boundary conditions if $h_2(x_1)\not=h_2(-x_1)$.

The matrix $\til{\Mcal}_\text{N}$ for the Neumann boundary condition
is obtained similarly as for the Dirichlet boundary condition.
Evaluating first the Fourier transform of the orthogonal components as
done
in expression (\ref{eq:TM-Matrix}), the result is
\begin{equation}\label{eq:TE-Matrix}
\begin{split}
\til{\Mcal}_\text{N}^{\alpha\beta}
\left(\pbf,\qbf\right)\:&=\:
(2\pi)^2\delta^{(2)}\left(\pbf_\perp\!+\qbf_\perp\right)\\[5pt]
&\qquad\times\,\int_{x_1}\!\int_{y_1}
e^{ip_1x_1+iq_1y_1}\,(-1)^{\alpha+\beta}
\left(-\pd_{x_3}^2+\left(h_\alpha'(x_1)+h_\beta'(y_1)\right)\,
\pd_{x_1}\pd_{x_3}-h_\alpha'(x_1)h_\beta'(y_1)\,\pd_{x_1}^2\right)\,
\\[5pt]
&\qquad\times\,\int_{p_1'}e^{-ip_1'(x_1-y_1)}\,
\frac{e^{-\sqrt{p_\perp^2\!+p_1'^2}\,
\abs{x_3-y_3}}}{2\sqrt{p_\perp^2+p_1'^2}}
\Big\vert_{\begin{smallmatrix}
x_3=h_\alpha(x_1)+H\delta_{\alpha2}\\y_3=h_\beta(y_1)+H\delta_{\beta2}
\end{smallmatrix}}
\\[5pt]
\:&=\:(2\pi)^2\delta^{(2)}(\pbf_\perp\!+\qbf_\perp)\int_{x_1}\int_{y_1}
\int_{p_1'}e^{i(p_1-p_1')x_1+i(q_1+p_1')y_1}\\[5pt]
&\qquad\times\,\frac{(-1)^{\alpha+\beta}}{2}\left[-\sqrt{p_\perp^2\!+p_1'^2}-
\frac{ip_1'}{\sqrt{p_\perp^2\!+p_1'^2}}\,
(\pd_{x_1}\!-\pd_{y_1})-
\frac{p_1'^2}{(p_\perp^2+p_1'^2)^{\frac{3}{2}}}\,
\pd_{x_1}\pd_{y_1}\right]
\\[5pt]
&\qquad\times\,
e^{-\sqrt{p_\perp^2\!+p_1'^2}\,
\abs{h_\alpha(x_1)-h_\beta(y_1)+H(\delta_{\alpha2}-\delta_{\beta2})}}.
\end{split}
\end{equation}
We apply partial integration to obtain
\begin{equation}\label{eq:TE-Matrix-2}
\begin{split}
\til{\Mcal}_\text{N}^{\alpha\beta}
\left(\pbf,\qbf\right)\:&=\:
(2\pi)^2\delta^{(2)}(\pbf_\perp\!+\qbf_\perp)
\\[5pt]
&\quad\times\,\frac{(-1)^{\alpha+\beta}}{2}\int_{p_1'}
\left[-\sqrt{p_\perp^2\!+p_1'^2}-
\frac{p_1'}{\sqrt{p_\perp^2\!+p_1'^2}}\,
\left(p_1-q_1-2p_1'\right)
+\,\frac{p_1'^2}{(p_\perp^2\!+p_1'^2)^{\frac{3}{2}}}(p_1-p_1')(q_1+p_1')
\right]\\[5pt]
&\quad\times\,\int_{x_1}\int_{y_1}
e^{i(p_1-p_1')x_1+i(q_1+p_1')y_1}\,
e^{-\sqrt{p_\perp^2\!+p_1'^2}\,
\abs{h_\alpha(x_1)-h_\beta(y_1)+H(\delta_{\alpha2}-\delta_{\beta2})}}.
\end{split}
\end{equation}

This expression will be treated analogously to the case of the matrix
for the Dirichlet boundary condition, cf. Eq.~(\ref{eq:TM-Matrix}).
It differs from the Dirichlet kernel by the additional $p_1'$
dependent term. This yields again Eq.~(\ref{eq:structure}), but now
with $N_{\text{D},m}$ substituted by the Neumann matrices
$N_{\text{N},m}$, which are given by
\begin{equation}
N_{\text{N},m}\left(q_\perp,q_1\right)\:=\:
\left(
\begin{array}{cc}
A^{\text{N}}_{m,1}\left(q_\perp,q_1\right) &
B^{\text{N}}_{m,12}\left(q_\perp,q_1\right)\\[5pt]
\gamma^mB^{\text{N}}_{m,21}\left(q_\perp,q_1\right) &
\gamma^mA^{\text{N}}_{m,2}\left(q_\perp,q_1\right)
\end{array}\right)\:+\:\delta_{m0}\left(\begin{array}{cc}
-\frac{q}{4}(1+e^{-2a_1q}) & \frac{q}{2}e^{-qH}\ch(a_1q)\ch(a_2q)\\[5pt]
\frac{q}{2}e^{-qH}\ch(a_1q)\ch(a_2q) & -\frac{q}{4}(1+e^{-2a_2q})
\end{array}\right)
\end{equation}
for $m$ even, and
\begin{equation}
N_{\text{N},m}\left(q_\perp,q_1\right)\:=\:
\left(
\begin{array}{cc}
0 & C^{\text{N}}_{m,12}\left(q_\perp,q_1\right)\\[5pt]
C^{\text{N}}_{m,21}\left(q_\perp,q_1\right) & 0
\end{array}\right)
\end{equation}
for $m$ odd. The entries are now given by
\begin{equation}
A^{\text{N}}_{m,\alpha}\left(q_\perp,q_1\right)\:=\:
\frac{(-1)^{\frac{m}{2}}}{\pi^2}\sum_{k=-\infty}^\infty
\frac{1}{(m-2k+1)(2k-1)}\,
\frac{1-e^{-2a_\alpha\til{q}_{2k-1}}}{\til{q}_{2k-1}^3}\,\phi_{mk}(q_\perp,q_1)\,,
\end{equation}
\begin{equation}
B^{\text{D}}_{m,\alpha\beta}\left(q_\perp,q_1\right)\:=\:
2\,\frac{(-1)^\frac{m}{2}}{\pi^2}\sum_{k=-\infty}^\infty\frac{
\gamma^{(2k-1)(\delta_{\beta2}-\delta_{\alpha2})}}
{(m-2k+1)(2k-1)}
\frac{e^{-\til{q}_{2k-1}H}}{\til{q}_{2k-1}^3}
\sh(a_\alpha\til{q}_{2k-1})\sh(a_\beta\til{q}_{2k-1})\,\phi_{mk}(q_\perp,q_1)\,,
\end{equation}
and
\begin{equation}
\begin{split}
C^{\text{N}}_{m,\alpha\beta}\left(q_\perp,q_1\right)\:&=\:
\frac{(-1)^\frac{m+1}{2}}{m\pi}
\bigg[
(-1)^\alpha\gamma^{m\delta_{\alpha2}}\,e^{-qH}
\left(q+\frac{2\pi m}{\lambda}\frac{q_1}{q}\right)
\sh(a_\alpha q)\ch(a_\beta q)\\[5pt]
&\qquad\qquad\qquad\quad+\:
(-1)^\beta\gamma^{m\delta_{\beta2}}
e^{-\til{q}_mH}\left(\til{q}_m-\frac{2\pi m}{\lambda}
\frac{q_1+2\pi m/\lambda}{\til{q}_m}\right)
\sh(a_\beta\til{q}_m)\ch(a_\alpha\til{q}_m)
\bigg]\,,
\end{split}
\end{equation}
using the function
\begin{equation}
\phi_{mk}(q_\perp,q_1)\:=\:q_1\left(q_1+\frac{2\pi m}{\lambda}\right)
\left(q_1+\frac{2\pi}{\lambda}(2k-1)\right)^2+2q_\perp^2
\left(q_1+\frac{\pi m}{\lambda}\right)\left(q_1+\frac{2\pi}{\lambda}
(2k-1)\right)+q_\perp^4\,.
\end{equation}
As in the case of the Dirichlet matrices, the off diagonal elements
depend on $b$ via the phase factor $\gamma=e^{\frac{2\pi
    i}{\lambda}b}$.  The matrices of the previous discussion of the
rectangular corrugation model are now simply recovered by performing
the limit $a_2\ra0$ and by defining $a=a_1$.


\section{The limit of small $\lambda$ for the matrices $N_m$}

In this section, the limit $\lambda\to0$ of the matrices
$N_m(q_\perp,q_1)$ for the rectangular corrugation model of section
III will be performed (cf. Appendix A for $a=a_1$, $a_2=0$ and
$\lambda\to0$).  These matrices depend on the shift of the argument
$q_1$ relative to $2\pi n/\lambda$ which requires a separate treatment
of various cases. Considering this, for the Dirichlet case we find the
simplified expressions
\begin{equation}
N_{\text{D},0}\left(q_\perp,q_1+2\pi n/\lambda\right)
\str{\lambda\ra0}{=}\left\{
\begin{array}{ll}
\left( \begin{array}{cc}
\frac{e^{-2aq}+1}{4q} &
\frac{e^{-qH}}{2q}\ch(aq)\\[5pt]
\frac{e^{-qH}}{2q}\ch(aq) & \frac{1}{2q}
\end{array}\right) & \mbox{for $n=0$}\\[20pt]
\left( \begin{array}{cc}
-\frac{1}{\pi^2n^2}\frac{e^{-2aq}-1}{q} & \epsilon\\[5pt]
\epsilon & \frac{\lambda}{4\pi\abs{n}}
\end{array}\right) & \mbox{for $n\,\text{odd}$}\\[20pt]
\left( \begin{array}{cc}
0 & \epsilon\\[5pt]
\epsilon & \frac{\lambda}{4\pi\abs{n}}
\end{array}\right) & \mbox{for $n\,\text{even}$}
\end{array}\right. .
\end{equation}
We have introduced a small quantity $\epsilon$, which is needed in
order to have a non singular matrix $B_{kl}$. However, at the end we
can safely take $\epsilon \to 0$ in the final expression for the
Casimir force. As $\lambda \to 0$, this quantity vanishes as
$\epsilon\sim\lambda\exp(-2\pi n(H-a)/\lambda)$.  The other matrices
for $m\not=0$ are given by
\begin{equation}
N_{\text{D},m}\left(q_\perp,q_1+2\pi n/\lambda\right)
\str{\lambda\ra0}{=}\left\{
\begin{array}{ll}
\left(\begin{array}{cc}
0 &
\frac{(-1)^\frac{m-1}{2}}{\pi m}\,\frac{e^{-qH}}{q}\sh(aq)\\[5pt]
0 & 0
\end{array}\right) & \mbox{for $n=0$}\\[20pt]
\left(\begin{array}{cc}
0 & 0\\[5pt]
\frac{(-1)^\frac{m-1}{2}}{\pi m}\,\frac{e^{-qH}}{q}\sh(aq) & 0
\end{array}\right) & \mbox{for $n=-m$}\\[20pt]
\left(\begin{array}{cc}
0 & 0\\[5pt]
0 & 0
\end{array}\right) & \mbox{for $n\not\in\{-m,0\}$}
\end{array}\right.
\end{equation}
for $m$ odd, and
\begin{equation}
N_{\text{D},m}\left(q_\perp,q_1+2\pi n/\lambda\right)
\str{\lambda\ra0}{=}\left\{
\begin{array}{ll}
\left(\begin{array}{cc}
-\frac{(-1)^\frac{m}{2}}{\pi^2n(m+n)}\,\frac{e^{-2aq}-1}{q} & 0\\[5pt]
0 & 0
\end{array}\right) & \mbox{for $n\,\text{odd}$}\\[20pt]
\left(\begin{array}{cc}
0 & 0\\[5pt]
0 & 0
\end{array}\right) & \mbox{for $n\,\text{even}$}
\end{array}\right.
\end{equation}
for even $m\not=0$. Analogously,
for the Neumann matrices, we find
\begin{equation}
N_{\text{N},0}\left(q_\perp,q_1+2\pi n/\lambda\right)
\str{\lambda\ra0}{=}
\left\{
\begin{array}{ll}
\left(\begin{array}{cc}
-\frac{q}{4}(e^{-2aq}+1) &
\frac{q}{2}e^{-qH}\ch(aq)\\[5pt]
\frac{q}{2}e^{-qH}\ch(aq) & -\frac{q}{2}
\end{array}\right) & \mbox{for $n=0$}\\[20pt]
\left(\begin{array}{cc}
\frac{4(-1)^{n-1}}{\lambda^2}\frac{q_1^2}{q^3}(e^{-2aq}-1)
& \epsilon\\[5pt]
\epsilon & -\frac{\pi\abs{n}}{\lambda}
\end{array}\right) & \mbox{for $n$ odd}\\[20pt]
\left( \begin{array}{cc}
-\frac{1}{\lambda}\big[\frac{\pi\abs{n}}{2}+
\frac{2}{\pi}\til{C}_0(n)\big] & \epsilon\\[5pt]
\epsilon & -\frac{\pi\abs{n}}{\lambda}
\end{array}\right) & \mbox{for $n$ even}
\end{array}\right.
\end{equation}
and
\begin{equation}
N_{\text{N},m}\left(q_\perp,q_1+2\pi n/\lambda\right)
\str{\lambda\ra0}{=}
\left\{
\begin{array}{ll}
\left(\begin{array}{cc}
0 &
\frac{2(-1)^\frac{m-1}{2}}{\lambda}\frac{q_1}{q}\,e^{-qH}\sh(aq)\\[5pt]
0 & 0
\end{array}\right) & \mbox{for $n=0$}\\[20pt]
\left(\begin{array}{cc}
0 & 0\\[5pt]
-\frac{2(-1)^\frac{m-1}{2}}{\lambda}\frac{q_1}{q}\,e^{-qH}\sh(aq) & 0
\end{array}\right) & \mbox{for $n=-m$}\\[20pt]
\left(\begin{array}{cc}
0 & 0\\[5pt]
0 & 0
\end{array}\right) & \mbox{for $n\not\in\{-m,0\}$}
\end{array}\right.
\end{equation}
for $m$ odd, and
\begin{equation}
N_{\text{N},m}\left(q_\perp,q_1+2\pi n/\lambda\right)
\str{\lambda\ra0}{=}\left\{
\begin{array}{ll}
\left(\begin{array}{cc}
\frac{4(-1)^\frac{m}{2}}{\lambda^2}\,\frac{q_1^2}{q^3}(e^{-2aq}-1) &
0\\[5pt]
0 & 0
\end{array}\right) & \mbox{for $n\,\text{odd}$}\\[20pt]
\left(\begin{array}{cc}
-\frac{2n(n+m)}{\pi\lambda}\,\til{C}_m(n) & 0\\[5pt]
0 & 0
\end{array}\right) & \mbox{for $n\,\text{even},n\not\in\{-m,0\}$}\\[20pt]
\left(\begin{array}{cc}
\pm\frac{mq_1}{\pi^2}\,\til{C}_m(n) & 0\\[5pt]
0 & 0
\end{array}\right) & \mbox{for $n\in\{-m,0\}$}
\end{array}\right.
\end{equation}
for even $m\not=0$. Here, the asymptotic behavior of $\epsilon$ for
$\lambda\ra0$ is $\epsilon\sim\lambda^{-1} \exp(-2\pi n(H\pm
a)/\lambda)$.  The constant is given by
$\til{C}_m(n)=(-1)^{m/2}\sum_{l=-\infty}'^\infty[(2l-1)(2l-1-m)\abs{2l-1+n}\,]^{-1}$,
and the prime at the summation sign indicates that $l\not=(1-n)/2$ if
$n$ is odd.

\acknowledgments

This work was supported by the Deutsche Forschungsgemeinschaft through
Emmy Noether grant No. EM70/2-2.


\begin{thebibliography}{38}
\expandafter\ifx\csname natexlab\endcsname\relax\def\natexlab#1{#1}\fi
\expandafter\ifx\csname bibnamefont\endcsname\relax
  \def\bibnamefont#1{#1}\fi
\expandafter\ifx\csname bibfnamefont\endcsname\relax
  \def\bibfnamefont#1{#1}\fi
\expandafter\ifx\csname citenamefont\endcsname\relax
  \def\citenamefont#1{#1}\fi
\expandafter\ifx\csname url\endcsname\relax
  \def\url#1{\texttt{#1}}\fi
\expandafter\ifx\csname urlprefix\endcsname\relax\def\urlprefix{URL }\fi
\providecommand{\bibinfo}[2]{#2}
\providecommand{\eprint}[2][]{\url{#2}}

\bibitem[{\citenamefont{Casimir}(1948)}]{H.B.G.Casimir}
\bibinfo{author}{\bibfnamefont{H.~B.~G.} \bibnamefont{Casimir}},
  \bibinfo{journal}{Proc. K. Ned. Akad. Wet.} \textbf{\bibinfo{volume}{51}},
  \bibinfo{pages}{793} (\bibinfo{year}{1948}).

\bibitem[{\citenamefont{Milloni}(1993)}]{milonni}
\bibinfo{author}{\bibfnamefont{P.~W.} \bibnamefont{Milloni}},
  \emph{\bibinfo{title}{The Quantum Vacuum}} (\bibinfo{publisher}{Acad. Press},
  \bibinfo{year}{1993}).

\bibitem[{\citenamefont{Bordag et~al.}(2001)\citenamefont{Bordag, Mohideen, and
  Mostepanenko}}]{Bordag-Mostepanenko}
\bibinfo{author}{\bibfnamefont{M.}~\bibnamefont{Bordag}},
  \bibinfo{author}{\bibfnamefont{U.}~\bibnamefont{Mohideen}}, \bibnamefont{and}
  \bibinfo{author}{\bibfnamefont{V.~M.} \bibnamefont{Mostepanenko}},
  \bibinfo{journal}{Phys. Rep.} \textbf{\bibinfo{volume}{353}},
  \bibinfo{pages}{1} (\bibinfo{year}{2001}).

\bibitem[{\citenamefont{Milton}(2001)}]{Milton01}
\bibinfo{author}{\bibfnamefont{K.~A.} \bibnamefont{Milton}},
  \emph{\bibinfo{title}{The Casimir effect: Physical Manifestations of
  Zero-Point Energy}} (\bibinfo{publisher}{World Scientific},
  \bibinfo{year}{2001}).

\bibitem[{\citenamefont{Kardar and Golestanian}(1999)}]{Kardar-Golestanian}
\bibinfo{author}{\bibfnamefont{M.}~\bibnamefont{Kardar}} \bibnamefont{and}
  \bibinfo{author}{\bibfnamefont{R.}~\bibnamefont{Golestanian}},
  \bibinfo{journal}{Rev. Mod. Phys.} \textbf{\bibinfo{volume}{71}},
  \bibinfo{pages}{1233} (\bibinfo{year}{1999}).

\bibitem[{\citenamefont{Israelachvili}(1992)}]{bwIsraelachvili}
\bibinfo{author}{\bibfnamefont{J.}~\bibnamefont{Israelachvili}},
  \emph{\bibinfo{title}{Intermolecular and Surface Forces}}
  (\bibinfo{publisher}{Academic Press, San Diego}, \bibinfo{year}{1992}).

\bibitem[{\citenamefont{Garcia and Chan}(2002)}]{Garcia-Chan}
\bibinfo{author}{\bibfnamefont{R.}~\bibnamefont{Garcia}} \bibnamefont{and}
  \bibinfo{author}{\bibfnamefont{M.~H.~W.} \bibnamefont{Chan}},
  \bibinfo{journal}{Phys. Rev. Lett.} \textbf{\bibinfo{volume}{88}},
  \bibinfo{pages}{086101} (\bibinfo{year}{2002}).

\bibitem[{\citenamefont{Bytsenko et~al.}(1996)\citenamefont{Bytsenko, Cognola,
  Vanzo, and Zerbini}}]{Bytsenko-Cognola-et-al}
\bibinfo{author}{\bibfnamefont{A.~A.} \bibnamefont{Bytsenko}},
  \bibinfo{author}{\bibfnamefont{G.}~\bibnamefont{Cognola}},
  \bibinfo{author}{\bibfnamefont{L.}~\bibnamefont{Vanzo}}, \bibnamefont{and}
  \bibinfo{author}{\bibfnamefont{S.}~\bibnamefont{Zerbini}},
  \bibinfo{journal}{Phys. Rep.} \textbf{\bibinfo{volume}{266}},
  \bibinfo{pages}{1} (\bibinfo{year}{1996}).

\bibitem[{\citenamefont{Milton}(1980{\natexlab{a}})}]{Milton80a}
\bibinfo{author}{\bibfnamefont{K.~A.} \bibnamefont{Milton}},
  \bibinfo{journal}{Phys. Rev. D} \textbf{\bibinfo{volume}{22}},
  \bibinfo{pages}{1441} (\bibinfo{year}{1980}{\natexlab{a}}).

\bibitem[{\citenamefont{Milton}(1980{\natexlab{b}})}]{Milton80b}
\bibinfo{author}{\bibfnamefont{K.~A.} \bibnamefont{Milton}},
  \bibinfo{journal}{Phys. Rev. D} \textbf{\bibinfo{volume}{22}},
  \bibinfo{pages}{1444} (\bibinfo{year}{1980}{\natexlab{b}}).

\bibitem[{\citenamefont{Lamoreaux}(1997)}]{Lamoreaux}
\bibinfo{author}{\bibfnamefont{S.~K.} \bibnamefont{Lamoreaux}},
  \bibinfo{journal}{Phys. Rev. Lett.} \textbf{\bibinfo{volume}{78}},
  \bibinfo{pages}{5} (\bibinfo{year}{1997}).

\bibitem[{\citenamefont{Mohideen and Roy}(1998)}]{Mohideen-Roy}
\bibinfo{author}{\bibfnamefont{U.}~\bibnamefont{Mohideen}} \bibnamefont{and}
  \bibinfo{author}{\bibfnamefont{A.}~\bibnamefont{Roy}},
  \bibinfo{journal}{Phys. Rev. Lett.} \textbf{\bibinfo{volume}{81}},
  \bibinfo{pages}{4549} (\bibinfo{year}{1998}).

\bibitem[{\citenamefont{Chan et~al.}(2001)\citenamefont{Chan, Aksyuk, Kleiman,
  Bishop, and Capasso}}]{Chan-et-al}
\bibinfo{author}{\bibfnamefont{H.~B.} \bibnamefont{Chan}},
  \bibinfo{author}{\bibfnamefont{V.~A.} \bibnamefont{Aksyuk}},
  \bibinfo{author}{\bibfnamefont{R.~N.} \bibnamefont{Kleiman}},
  \bibinfo{author}{\bibfnamefont{D.~J.} \bibnamefont{Bishop}},
  \bibnamefont{and} \bibinfo{author}{\bibfnamefont{F.}~\bibnamefont{Capasso}},
  \bibinfo{journal}{Science} \textbf{\bibinfo{volume}{291}},
  \bibinfo{pages}{1941} (\bibinfo{year}{2001}).

\bibitem[{\citenamefont{Bressi et~al.}(2002)\citenamefont{Bressi, Carugno,
  Onofrio, and Ruoso}}]{Bressi-et-al}
\bibinfo{author}{\bibfnamefont{G.}~\bibnamefont{Bressi}},
  \bibinfo{author}{\bibfnamefont{G.}~\bibnamefont{Carugno}},
  \bibinfo{author}{\bibfnamefont{R.}~\bibnamefont{Onofrio}}, \bibnamefont{and}
  \bibinfo{author}{\bibfnamefont{G.}~\bibnamefont{Ruoso}},
  \bibinfo{journal}{Phys. Rev. Lett.} \textbf{\bibinfo{volume}{88}},
  \bibinfo{pages}{041804} (\bibinfo{year}{2002}).

\bibitem[{\citenamefont{Derjaguin}(1934)}]{Derjaguin}
\bibinfo{author}{\bibfnamefont{B.}~\bibnamefont{Derjaguin}},
  \bibinfo{journal}{Kolloid Z.} \textbf{\bibinfo{volume}{69}},
  \bibinfo{pages}{155} (\bibinfo{year}{1934}).

\bibitem[{\citenamefont{Boyer}(1974)}]{Boyer}
\bibinfo{author}{\bibfnamefont{T.}~\bibnamefont{Boyer}},
  \bibinfo{journal}{Phys. Rev. A} \textbf{\bibinfo{volume}{9}},
  \bibinfo{pages}{2078} (\bibinfo{year}{1974}).

\bibitem[{\citenamefont{Iannuzzi et~al.}()\citenamefont{Iannuzzi, Gelfand,
  Lisanti, and Capasso}}]{Iannuzzi+03}
\bibinfo{author}{\bibfnamefont{D.}~\bibnamefont{Iannuzzi}},
  \bibinfo{author}{\bibfnamefont{I.}~\bibnamefont{Gelfand}},
  \bibinfo{author}{\bibfnamefont{M.}~\bibnamefont{Lisanti}}, \bibnamefont{and}
  \bibinfo{author}{\bibfnamefont{F.}~\bibnamefont{Capasso}},
  \bibinfo{note}{preprint quant-ph/0312043}.

\bibitem[{\citenamefont{Maclay}(2000)}]{Maclay00}
\bibinfo{author}{\bibfnamefont{G.~J.} \bibnamefont{Maclay}},
  \bibinfo{journal}{Phys. Rev. A} \textbf{\bibinfo{volume}{61}},
  \bibinfo{pages}{052110} (\bibinfo{year}{2000}).

\bibitem[{\citenamefont{Serry et~al.}(1995)\citenamefont{Serry, Walliser, and
  Maclay}}]{Serry-Walliser}
\bibinfo{author}{\bibfnamefont{F.~M.} \bibnamefont{Serry}},
  \bibinfo{author}{\bibfnamefont{D.}~\bibnamefont{Walliser}}, \bibnamefont{and}
  \bibinfo{author}{\bibfnamefont{G.~J.} \bibnamefont{Maclay}},
  \bibinfo{journal}{J. Microelectromech. Syst.} \textbf{\bibinfo{volume}{4}},
  \bibinfo{pages}{193} (\bibinfo{year}{1995}).

\bibitem[{\citenamefont{Roy and Mohideen}(1999)}]{Roy-Mohideen}
\bibinfo{author}{\bibfnamefont{A.}~\bibnamefont{Roy}} \bibnamefont{and}
  \bibinfo{author}{\bibfnamefont{U.}~\bibnamefont{Mohideen}},
  \bibinfo{journal}{Phys. Rev. Lett.} \textbf{\bibinfo{volume}{82}},
  \bibinfo{pages}{4380} (\bibinfo{year}{1999}).

\bibitem[{\citenamefont{Chen et~al.}(2002)\citenamefont{Chen, Mohideen,
  Klimchitskaya, and Mostepanenko}}]{Chen-Mohideen-Klim-Mostepenner}
\bibinfo{author}{\bibfnamefont{F.}~\bibnamefont{Chen}},
  \bibinfo{author}{\bibfnamefont{U.}~\bibnamefont{Mohideen}},
  \bibinfo{author}{\bibfnamefont{G.~L.} \bibnamefont{Klimchitskaya}},
  \bibnamefont{and} \bibinfo{author}{\bibfnamefont{V.~M.}
  \bibnamefont{Mostepanenko}}, \bibinfo{journal}{Phys. Rev. Lett.}
  \textbf{\bibinfo{volume}{88}}, \bibinfo{pages}{101801}
  (\bibinfo{year}{2002}).

\bibitem[{\citenamefont{Klimchitskaya et~al.}(2001)\citenamefont{Klimchitskaya,
  Zanette, and Caride}}]{Klim-Zanette-Caride}
\bibinfo{author}{\bibfnamefont{G.~L.} \bibnamefont{Klimchitskaya}},
  \bibinfo{author}{\bibfnamefont{S.~I.} \bibnamefont{Zanette}},
  \bibnamefont{and} \bibinfo{author}{\bibfnamefont{O.~A.}
  \bibnamefont{Caride}}, \bibinfo{journal}{Phys. Rev. A}
  \textbf{\bibinfo{volume}{63}}, \bibinfo{pages}{014101}
  (\bibinfo{year}{2001}).

\bibitem[{\citenamefont{Golestanian and
  Kardar}(1998{\natexlab{a}})}]{golkar-letter}
\bibinfo{author}{\bibfnamefont{R.}~\bibnamefont{Golestanian}} \bibnamefont{and}
  \bibinfo{author}{\bibfnamefont{M.}~\bibnamefont{Kardar}},
  \bibinfo{journal}{Phys. Rev. Lett.} \textbf{\bibinfo{volume}{78}},
  \bibinfo{pages}{3421} (\bibinfo{year}{1998}{\natexlab{a}}).

\bibitem[{\citenamefont{Golestanian and Kardar}(1998{\natexlab{b}})}]{golkar}
\bibinfo{author}{\bibfnamefont{R.}~\bibnamefont{Golestanian}} \bibnamefont{and}
  \bibinfo{author}{\bibfnamefont{M.}~\bibnamefont{Kardar}},
  \bibinfo{journal}{Phys. Rev. A} \textbf{\bibinfo{volume}{58}},
  \bibinfo{pages}{1713} (\bibinfo{year}{1998}{\natexlab{b}}).

\bibitem[{\citenamefont{Emig et~al.}(2001)\citenamefont{Emig, Hanke,
  Golestanian, and Kardar}}]{Emig-Hanke-Golestanian-Kardar}
\bibinfo{author}{\bibfnamefont{T.}~\bibnamefont{Emig}},
  \bibinfo{author}{\bibfnamefont{A.}~\bibnamefont{Hanke}},
  \bibinfo{author}{\bibfnamefont{R.}~\bibnamefont{Golestanian}},
  \bibnamefont{and} \bibinfo{author}{\bibfnamefont{M.}~\bibnamefont{Kardar}},
  \bibinfo{journal}{Phys. Rev. Lett.} \textbf{\bibinfo{volume}{87}},
  \bibinfo{pages}{260402} (\bibinfo{year}{2001}).

\bibitem[{\citenamefont{Emig et~al.}(2003)\citenamefont{Emig, Hanke,
  Golestanian, and Kardar}}]{Emig-Hanke-Golestanian-Kardar-II}
\bibinfo{author}{\bibfnamefont{T.}~\bibnamefont{Emig}},
  \bibinfo{author}{\bibfnamefont{A.}~\bibnamefont{Hanke}},
  \bibinfo{author}{\bibfnamefont{R.}~\bibnamefont{Golestanian}},
  \bibnamefont{and} \bibinfo{author}{\bibfnamefont{M.}~\bibnamefont{Kardar}},
  \bibinfo{journal}{Phys. Rev. A} \textbf{\bibinfo{volume}{67}},
  \bibinfo{pages}{022114} (\bibinfo{year}{2003}).

\bibitem[{\citenamefont{Balian and Duplantier}(1978)}]{Balian-Duplantier}
\bibinfo{author}{\bibfnamefont{R.}~\bibnamefont{Balian}} \bibnamefont{and}
  \bibinfo{author}{\bibfnamefont{B.}~\bibnamefont{Duplantier}},
  \bibinfo{journal}{Ann. Phys. (N.Y.)} \textbf{\bibinfo{volume}{112}},
  \bibinfo{pages}{165} (\bibinfo{year}{1978}).

\bibitem[{\citenamefont{Jaffe and Scardicchio}()}]{jaffe}
\bibinfo{author}{\bibfnamefont{R.~L.} \bibnamefont{Jaffe}} \bibnamefont{and}
  \bibinfo{author}{\bibfnamefont{A.}~\bibnamefont{Scardicchio}},
  \bibinfo{note}{preprint quant-ph/0310194}.

\bibitem[{\citenamefont{Li and Kardar}(1991)}]{likar-letter}
\bibinfo{author}{\bibfnamefont{H.}~\bibnamefont{Li}} \bibnamefont{and}
  \bibinfo{author}{\bibfnamefont{M.}~\bibnamefont{Kardar}},
  \bibinfo{journal}{Phys. Rev. Lett.} \textbf{\bibinfo{volume}{67}},
  \bibinfo{pages}{3257} (\bibinfo{year}{1991}).

\bibitem[{\citenamefont{Li and Kardar}(1992)}]{likar}
\bibinfo{author}{\bibfnamefont{H.}~\bibnamefont{Li}} \bibnamefont{and}
  \bibinfo{author}{\bibfnamefont{M.}~\bibnamefont{Kardar}},
  \bibinfo{journal}{Phys. Rev. A} \textbf{\bibinfo{volume}{46}},
  \bibinfo{pages}{6490} (\bibinfo{year}{1992}).

\bibitem[{\citenamefont{Emig}(2003)}]{emig}
\bibinfo{author}{\bibfnamefont{T.}~\bibnamefont{Emig}},
  \bibinfo{journal}{Europhys. Lett.} \textbf{\bibinfo{volume}{62}},
  \bibinfo{pages}{466} (\bibinfo{year}{2003}).

\bibitem[{\citenamefont{Peskin and Schroeder}(1995)}]{peskin}
\bibinfo{author}{\bibfnamefont{M.~E.} \bibnamefont{Peskin}} \bibnamefont{and}
  \bibinfo{author}{\bibfnamefont{D.~V.} \bibnamefont{Schroeder}},
  \emph{\bibinfo{title}{An Introduction to Quantum Field Theory}}
  (\bibinfo{publisher}{Addison-Wesley, Reading, MA}, \bibinfo{year}{1995}).

\bibitem[{\citenamefont{Jackson}(1967)}]{jackson}
\bibinfo{author}{\bibfnamefont{J.~D.} \bibnamefont{Jackson}},
  \emph{\bibinfo{title}{Classical Electrodynamics}} (\bibinfo{publisher}{John
  Wiley \& Sons, Inc.}, \bibinfo{year}{1967}).

\bibitem[{\citenamefont{Karepanov et~al.}(1987)\citenamefont{Karepanov,
  Novikov, and Sorin}}]{Karepanov+87}
\bibinfo{author}{\bibfnamefont{S.~K.} \bibnamefont{Karepanov}},
  \bibinfo{author}{\bibfnamefont{M.~Y.} \bibnamefont{Novikov}},
  \bibnamefont{and} \bibinfo{author}{\bibfnamefont{A.~S.} \bibnamefont{Sorin}},
  \bibinfo{journal}{Nuovo Cimento Soc. Ital. Fis., B}
  \textbf{\bibinfo{volume}{100}}, \bibinfo{pages}{411} (\bibinfo{year}{1987}).

\bibitem[{\citenamefont{Klimchitskaya and Pavlov}(1996)}]{Klimchitskaya+96}
\bibinfo{author}{\bibfnamefont{G.~L.} \bibnamefont{Klimchitskaya}}
  \bibnamefont{and} \bibinfo{author}{\bibfnamefont{Y.~V.}
  \bibnamefont{Pavlov}}, \bibinfo{journal}{Int. J. Mod. Phys. A}
  \textbf{\bibinfo{volume}{11}}, \bibinfo{pages}{3723} (\bibinfo{year}{1996}).

\bibitem[{\citenamefont{Mostepanenko and Trunov}(1988)}]{Mostepanenko+88}
\bibinfo{author}{\bibfnamefont{V.~M.} \bibnamefont{Mostepanenko}}
  \bibnamefont{and} \bibinfo{author}{\bibfnamefont{N.~N.}
  \bibnamefont{Trunov}}, \bibinfo{journal}{Sov. Phys. Usp.}
  \textbf{\bibinfo{volume}{31}}, \bibinfo{pages}{965} (\bibinfo{year}{1988}).

\bibitem[{\citenamefont{Gies et~al.}(2003)\citenamefont{Gies, Langfeld, and
  Moyaerts}}]{Gies+03}
\bibinfo{author}{\bibfnamefont{H.}~\bibnamefont{Gies}},
  \bibinfo{author}{\bibfnamefont{K.}~\bibnamefont{Langfeld}}, \bibnamefont{and}
  \bibinfo{author}{\bibfnamefont{L.}~\bibnamefont{Moyaerts}},
  \bibinfo{journal}{JHEP} \textbf{\bibinfo{volume}{06}}, \bibinfo{pages}{018}
  (\bibinfo{year}{2003}).

\bibitem[{\citenamefont{B\"uscher and Emig}()}]{Buescher+03}
\bibinfo{author}{\bibfnamefont{R.}~\bibnamefont{B\"uscher}} \bibnamefont{and}
  \bibinfo{author}{\bibfnamefont{T.}~\bibnamefont{Emig}},
  \bibinfo{note}{preprint cond-mat/0308412}.

\end{thebibliography}
\end{document}